\documentclass[abstract,headings=normal,DIV=12]{scrartcl}

\usepackage[automark]{scrpage2}
\usepackage[english]{babel}
\usepackage[T1]{fontenc}
\usepackage[applemac]{inputenc}
\usepackage{graphicx}
\usepackage{tikz}
\usepackage{stmaryrd}
\usepackage{amsmath}
\usepackage{amssymb}
\usepackage{amsthm}
\usepackage{subfig}
\usepackage{accents}
\usepackage{url}
\usepackage{booktabs}
\usepackage[colorlinks=true,linkcolor=blue,citecolor=red,urlcolor=black]{hyperref}

\usetikzlibrary{arrows}

\addtokomafont{caption}{\small}

\newcommand{\R}{\mathbb{R}}
\newcommand{\Z}{\mathbb{Z}}

\newcommand{\Ell}{\mathcal{L}}
\newcommand{\ELL}{\mathfrak{L}}

\newcommand{\Es}{\mathcal{S}}
\newcommand{\E}{\mathcal{E}}

\newcommand{\X}{\mathcal{X}}
\newcommand{\N}{\mathbb{N}}

\DeclareMathOperator{\grad}{grad}
\DeclareMathOperator{\Li}{Li}

\renewcommand{\Re}{\operatorname{Re}}

\newcommand{\unbar}[1]{\underaccent{\bar}{#1}}
\renewcommand{\tilde}[1]{\widetilde{#1}}

\makeatletter
\renewcommand*\env@cases[1][1.2]{%
  \let\@ifnextchar\new@ifnextchar
  \left\lbrace
  \def\arraystretch{#1}%
  \array{@{}l@{\quad}l@{}}%
}
\makeatother

\newtheorem{theo}{Theorem}[section]
\newtheorem{lemma}[theo]{Lemma}

\theoremstyle{definition}
\newtheorem{defi}[theo]{Definition}
\theoremstyle{remark}

\graphicspath{{./graphics/}}
\DeclareGraphicsExtensions{.pdf}

\begin{document}

\title{On the variational interpretation\\ of the discrete KP equation}
\author{Raphael~Boll\and Matteo~Petrera\and Yuri~B.~Suris}
\publishers{\vspace{0.5cm}{\small Institut f\"ur Mathematik, MA 7-2, Technische Universit\"at Berlin,\\
Stra{\ss}e des 17. Juni 136, 10623 Berlin, Germany\\
E-mail: \url{boll}, \url{petrera}, \url{suris@math.tu-berlin.de}}}
\maketitle

\begin{abstract}
We study the variational structure of the discrete Kadomtsev-Petviashvili (dKP) equation by means of its pluri-Lagrangian formulation. We consider the dKP equation and its variational formulation on the cubic lattice $\Z^{N}$ as well as on the root lattice $Q(A_{N})$. We prove that, on a lattice of dimension at least four, the corresponding Euler-Lagrange equations are equivalent to the dKP equation.
\end{abstract}

\section{Introduction}\label{sec:intro}
We developed the theory of pluri-Lagrangian problems (integrable systems of variational origin) in recent papers \cite{S12,nonrel,S13,variational,BS2,rel,octahedron}, influenced by the fundamental insight of \cite{LN,LNQ,LN2,YLN}.
In the present paper, we consider the pluri-Lagrangian formulation of the discrete Kadomtsev-Petviashvili (dKP) equation on three-dimensional lattices and its consistent extension to higher dimensional lattices. This equation belongs to integrable octahedron-type equations which were classified in~\cite{ABS3}. A Lagrangian formulation of this equation was given in~\cite{LNQ}. There, the authors consider a discrete 3-form on the lattice $\Z^{3}$ together with the corresponding Euler-Lagrange equations which are shown to be satisfied on solutions of the dKP equation. They also show that this 3-form is closed on solutions of the dKP equation, namely, the so-called 4D~closure relation is satisfied. The main goal of the present paper is to provide a more precise understanding of the findings in that paper. More concretely:
\begin{itemize}
\item In the framework of the pluri-Lagrangian formulation, we construct the elementary building blocks of Euler-Lagrange equations, which, in the present situation, are the so-called 4D~corner equations.
\item In the two-dimensional case, as noticed in~\cite{variational}, the corresponding 3D~corner equations build a consistent system. Its solutions are more general then the solutions of the underlying hyperbolic system of quad-equations. On the contrary, in the present three-dimensional situation, the system of 4D~corner equations is not consistent in the usual sense (i.e., it does not allow to determine general solutions with the maximal number of initial data). However, this system turns out to be equivalent, in a sense which we are going to explain later, to the corresponding hyperbolic system, namely the dKP equation.
\item We provide a rigorous consideration of the branches of the logarithm functions involved in the Euler-Lagrange equations. This leads to the following more precise result: the system of 4D~corner equations is equivalent, and thus provides a variational formulation, to two different hyperbolic equations, namely the dKP equation itself and its version obtained under inversion $x\mapsto x^{-1}$ of all fields which will be denoted by $\mathrm{dKP}^{-}$.
\end{itemize}\par
One can consider the dKP equation on the cubic lattice $\Z^{3}$ and its higher dimensional analogues $\Z^{N}$, but, as discussed in~\cite{ABS3} another natural setting the dKP equation (and related octahedron-type equations) is the three-dimensional root lattice
\[
Q(A_{3}):=\{(n_{i},n_{j},n_{k},n_{\ell}):n_{i}+n_{j}+n_{k}+n_{\ell}=0\}.
\]
Also in this setting, the dKP equation can be extended in a consistent way to the higher dimensional lattices $Q(A_{N})$ with $N>3$.

Both lattices have their advantages and disadvantages. The cubic lattice $\Z^{N}$, on the one hand, is more manageable and easier to visualize. Its cell structure is very simple: for every dimension $N$, all $N$-dimensional elementary cells are $N$-dimensional cubes. On the other hand, it is less natural to consider dKP on the lattice $\Z^{3}$, because this equation depends on the variables assigned to six out of eight vertices of a (three-dimensional) cube.

The root lattice $Q(A_{N})$, in contrast, has a more complicated cell structure, because the number of different $N$-dimensional elementary cells increases with the dimension $N$. For instance, for $N=3$ there are two types of elementary cells octahedra and tetrahedra. Moreover, especially in higher dimensions, a visualization of the elementary cells is difficult, if not impossible. However, this lattice is more natural for the consideration of dKP from the combinatorial point of view, because this equation depends on variables which can be assigned to the six vertices of an octahedron, one of the elementary cells of the lattice. Furthermore, the four-dimensional elementary cells are combinatorially smaller (they contain only 10 vertices, as compared with 16 vertices of a four-dimensional cube) and possess higher symmetry than the cubic ones. Since they support the equations which serve as variational analogue of the dKP equation, this leads to a simpler situation.

We will see that a four-dimensional cube is combinatorially equivalent to the sum of four elementary cells of the root lattice $Q(A_{4})$. Therefore, several results in the cubic case can be seen as direct consequences of results of the more fundamental $Q(A_{N})$-case.\par\medskip
Let us start with some concrete definitions valid for an arbitrary $N$-dimensional lattice $\X$.
\begin{defi}[Discrete 3-form]
A \emph{discrete 3-form} on $\X$ is a real-valued function $\Ell$ of oriented 3-cells~$\sigma$ depending on some field $x:\X\to \R$, such that $\Ell$ changes the sign by changing the orientation of~$\sigma$.
\end{defi}
For instance, in $Q(A_{N})$, the 3-cells are tetrahedra and octahedra, and, in $\Z^{N}$, the 3-cells are 3D cubes.
\begin{defi}[3-dimensional pluri-Lagrangian problem]
Let $\Ell$ be a discrete 3-form on $\X$ depending on $x:\X\to\R$.
\begin{itemize}
\item To an arbitrary 3-manifold $\Sigma\subset\X$, i.e., a union of oriented 3-cells which forms an oriented three-dimensional topological manifold, there corresponds the \emph{action functional}, which assigns to $x|_{V(\Sigma)}$, i.e., to the fields in the set of the vertices $V(\Sigma)$ of $\Sigma$, the number
\begin{equation*}
S_{\Sigma}:=\sum_{\sigma\in\Sigma}\Ell(\sigma).
\end{equation*}
\item We say that the field $x:V(\Sigma)\to \R$ is a critical point of $S_{\Sigma}$, if at any interior point $n\in V(\Sigma)$, we have
\begin{equation}\label{eq: dEL gen}
\frac{\partial S_{\Sigma}}{\partial x(n)}=0.
\end{equation}
Equations \eqref{eq: dEL gen} are called \emph{discrete Euler-Lagrange equations} for the action $S_{\Sigma}$.
\item We say that the field $x:\X\to\R$ solves the \emph{pluri-Lagrangian problem} for the Lagrangian 3-form $\Ell$ if, \emph{for any 3-manifold $\Sigma\subset\X$}, the restriction $x|_{V(\Sigma)}$ is a critical point of the corresponding action $S_{\Sigma}$.
\end{itemize}
\end{defi}
In the present paper, we focus on the variational formulation of the dKP equation on $Q(A_{N})$ and $\Z^{N}$. Let us formulate the main results of the paper.\par
On the lattice $Q(A_{N})$, we consider discrete 3-forms vanishing on all tetrahedra. One can show (see Corollary~\ref{cor:Corner}) that, for an arbitrary interior vertex of any 3-manifold in $Q(A_{N})$, the Euler-Lagrange equations follow from certain elementary building blocks. These so-called 4D~corner equations are the Euler-Lagrange equations for elementary 4-cells of $Q(A_{N})$ different from 4-simplices, so-called 4-ambo-simplices. Such a 4-ambo-simplex has ten vertices. Therefore, the crucial issue is the study of the system consisting of the corresponding ten corner equations. In our case, each corner equation depends on all ten fields at the vertices of the 4-ambo-simplex. Therefore, one could call this system \emph{consistent} if any two equations are functionally dependent. It turns out that this is \emph{not} the case. We will prove the following statement:
\begin{theo}\label{th:equiv}
Every solution of the system of ten corner equations for a 4-ambo-simplex in $Q(A_{N})$ satisfies either the system of five dKP equations or the system of five $\mathit{dKP}^{-}$ equations on the five octahedral facets of the 4-ambo-simplex.
\end{theo}
Thus, one can prescribe arbitrary initial values at seven vertices of a 4-ambo-simplex. We will also prove the following theorem:
\begin{theo}\label{th:clos}
The discrete 3-form $\Ell$ is closed on any solution of the system of corner equations.
\end{theo}
In~\cite{S12,variational}, it was shown that in dimensions 1 and 2 the analogues of the property formulated in Theorem~\ref{th:clos} are related to more traditional integrability attributes.

For the case of the cubic lattice $\Z^{N}$, the situation is similar: one can show (see Corollary~\ref{cor:Corner cubic}) that, for an arbitrary interior vertex of any 3-manifold in $\Z^{3}$, the Euler-Lagrange equations follow from certain elementary building blocks. These so-called 4D~corner equations are the Euler-Lagrange equations for elementary 4D~cubes in $\Z^{N}$. A 4D~cube has sixteen vertices, but in our case the action on a 4D~cube turns out to be independent of the fields on two of the vertices. Therefore, the crucial issue is the study of the system consisting of the corresponding fourteen corner equations. Six of the fourteen corner equations depend each on thirteen of the fourteen fields. There do not exist pairs of such equations which are independent of one and the same field. All other equations depend each on ten of the fourteen fields. Therefore, one could call this system \emph{consistent} if it would have the minimal possible rank 2 (assign twelve fields arbitrarily and use two of the six corner equations -- depending on thirteen fields -- to determine the remaining two fields, then all twelve remaining equations should be satisfied automatically). It turns out that the system of the fourteen corner equations is \emph{not} consistent in this sense. We will prove the following analogue of Theorem~\ref{th:equiv}:
\begin{theo}
Every solution of the system of fourteen corner equations for a 4D~cube in $\Z^{N}$ satisfies either the system of eight dKP equations or the system of eight $\mathit{dKP}^{-}$ equations on the eight cubic facets of the 4D~cube.
\end{theo}
Thus, one can prescribe arbitrary initial values at nine vertices of a 4D~cube. Correspondingly, we will also prove the following statement:
\begin{theo}
The discrete 3-form $\Ell$ is closed on any solution of the system of corner equations.
\end{theo}
The paper is organized as follows: we start with the root lattice $Q(A_{N})$, thus considering the combinatorial issues and some general properties of pluri-Lagrangian systems. Then we introduce the dKP equation and its pluri-Lagrangian structure. In the second part of the paper the present similar considerations for the cubic lattice $\Z^{N}$.

\section{The root lattice \texorpdfstring{$Q(A_{N})$}{Q(A\_N)}}\label{sect:root}
We consider the root lattice
\[
Q(A_{N}):=\{n:=(n_{0},n_{1},\ldots,n_{N})\in\Z^{N+1}:n_{0}+n_{1}+\ldots+n_{N}=0\},
\]
where $N\geq3$. The three-dimensional sub-lattices $Q(A_{3})$ are given by
\[
Q(A_{3}):=\{(n_{i},n_{j},n_{k},n_{\ell}):n_{i}+n_{j}+n_{k}+n_{\ell}=\mathrm{const}\}.
\]\par
We consider fields $x:Q(A_{N})\to\R$, and use the shorthand notations
\[
x_{\bar{\imath}}=x(n-e_{i}),\qquad x=x(n),\qquad\text{and}\qquad x_{i}=x(n+e_{i}),
\]
where $e_{i}$ is the unit vector in the $i$\textsuperscript{th} coordinate direction.
Furthermore, the shift functions $T_{i}$ and $T_{\bar{\imath}}$ are defined by
\[
T_{i}x_{\alpha}:=x_{i\alpha}\quad\text{and}\quad T_{\bar{\imath}}x_{\alpha}:=x_{\bar{\imath}\alpha}
\]
for a multiindex $\alpha$. For simplicity, we sometimes abuse notations by identifying lattice points $n$ with the corresponding fields $x(n)$.\par
We now give a very brief introduction to the Delaunay cell structure of the $n$-dimensional root lattice $Q(A_{N})$~\cite{CS,MP}. Here, we restrict ourselves to a very elementary description which is appropriate to our purposes and follow the considerations in \cite{ABS3}. For each $N$ there are $N$ sorts of $N$-cells of $Q(A_{N})$ denoted by $P(k,N)$ with $k=1,\ldots,N$:
\begin{itemize}
\item Two sorts of 2-cells:\ \\
\begin{tabular}{ll}\addlinespace
$P(1,2)$:& black triangles $\lfloor ijk\rfloor:=\{x_{i},x_{j},x_{k}\}$;\\\addlinespace
$P(2,2)$:& white triangles $\lceil ijk\rceil:=\{x_{ij},x_{ik},x_{jk}\}$;
\end{tabular}
\item Three sorts of 3-cells:\ \\
\begin{tabular}{ll}\addlinespace
$P(1,3)$:& black tetrahedra $\lfloor ijk\ell\rfloor:=\{x_{i},x_{j},x_{k},x_{\ell }\}$;\\\addlinespace
$P(2,3)$:& octahedra $[ijk\ell]:=\{x_{ij},x_{ik},x_{i\ell},x_{jk},x_{j\ell},x_{k\ell}\}$;\\\addlinespace
$P(3,3)$:& white tetrahedra $\lceil ijk\ell\rceil:=\{x_{ijk},x_{ij\ell},x_{ik\ell},x_{jk\ell}\}$;
\end{tabular}
\item Four sorts of 4-cells:\ \\
\begin{tabular}{ll}\addlinespace
$P(1,4):$ & black 4-simplices $\llfloor ijk\ell m\rrfloor:=\{x_{i},x_{j},x_{k},x_{\ell},x_{m}\}$;\\\addlinespace
$P(2,4):$ & black 4-ambo-simplices $\lfloor ijk\ell m\rfloor:=\{x_{\alpha\beta}:\alpha,\beta\in\{i,j,k,\ell,m\},\alpha\neq\beta\}$;\\\addlinespace
$P(3,4):$ & white 4-ambo-simplices $\lceil ijk\ell m\rceil:=\{x_{\alpha\beta\gamma}:\alpha,\beta,\gamma\in\{i,j,k,\ell,m\},$\\
& \hfill $\alpha\neq\beta\neq\gamma\neq\alpha\}$;\\\addlinespace
$P(4,4):$ & white 4-simplices $\llceil ijk\ell m\rrceil:=\{x_{ijk\ell},x_{ijkm},x_{ij\ell m},x_{ik\ell m},x_{jk\ell m}\}$.
\end{tabular}\bigskip
\end{itemize}
The facets of $3$-cells and $4$-cells can be found in Appendix~\ref{sec:facets}.\par
In the present paper we will consider objects on \emph{oriented} manifolds. We say that a black triangle $\lfloor ijk\rfloor$ and white triangle $\lceil ijk\rceil$ are positively oriented if $i<j<k$ (see Figure~\ref{fig:orientation}). Any permutation of two indices changes the orientation to the opposite one.\par
\begin{figure}[htbp]
   \centering
   \subfloat[]{\label{fig:black triangle}
   \begin{tikzpicture}[scale=0.85,inner sep=2]
      \node (x) at (0,0) [circle,fill,label=-135:$x_{i}$] {};
      \node (x1) at (3,0) [circle,fill,label=-45:$x_{j}$] {};
      \node (x3) at (0,3) [circle,fill,label=135:$x_{k}$] {};
      \draw (x) to (x1);
      \draw (x) to (x3);
      \draw (x1) to (x3);
      \draw [ultra thick,->] (0.38,0.88) arc (180:510:0.5);
   \end{tikzpicture}
   }\qquad
   \subfloat[]{\label{fig:white triangle}
      \begin{tikzpicture}[scale=0.85,inner sep=2]
      \node (x12) at (4,1) [circle,fill,label=-45:$x_{ij}$] {};
      \node (x23) at (1,4) [circle,fill,label=135:$x_{jk}$] {};
      \node (x123) at (4,4) [circle,fill,label=45:$x_{ik}$] {};
      \draw (x12) to (x123);
      \draw (x12) to (x23);
      \draw (x23) to (x123);
      \draw [ultra thick,->] (2.62,3.12) arc (180:510:0.5);
   \end{tikzpicture}
   }
   \caption{Orientation of triangles: \protect\subref{fig:black triangle} the black triangle $\lfloor ijk\rfloor$; \protect\subref{fig:white triangle} the white triangle $\lceil ijk\rceil$}
   \label{fig:orientation}
\end{figure}
When we use the bracket notation, we always write the letters in brackets in increasing order, so, e.g., in writing $\lfloor ijk\rfloor$ we assume that $i<j<k$ and avoid the notation $\lfloor jik\rfloor$ or $\lfloor ikj\rfloor$ for the negatively oriented triangle $-\lfloor ijk\rfloor$.\par
There is a simple recipe to derive the orientation of facets of an $N$-cell: On every index in the brackets we put alternately a ``$+$'' or a ``$-$'' starting with a ``$+$'' on the last index. Then we get each of its facets by deleting one index and putting the corresponding sign in front of the bracket. For instance, the black 4-ambo-simplex
{
\setlength\arraycolsep{0pt}
\renewcommand{\arraystretch}{0.5}
\begin{equation*}
\begin{array}{ccccccc}
&+&-&+&-&+&\\
\lfloor&i&j&k&\ell&m&\rfloor
\end{array}
\end{equation*}
has the five octahedral facets} $[ijk\ell]$, $-[ijkm]$, $[ij\ell m]$, $-[ik\ell m]$, and $[jk\ell m]$.
\par
The following two definitions are valid for arbitrary $N$-dimensional lattices $\X$.
\begin{defi}[Adjacent $N$-cell]
Given an $N$-cell $\sigma$, another $N$-cell $\bar{\sigma}$ is called \emph{adjacent} to $\sigma$ if $\sigma$ and $\bar{\sigma}$ share a common $(N-1)$-cell. The orientation of this $(N-1)$-cell in $\sigma$ must be opposite to its orientation in $\bar{\sigma}$.
\end{defi}
The latter property guarantees that the orientations of the adjacent $N$-cells agree.
\begin{defi}[Flower]\label{def:flower}
A 3-manifold in $\X$ with exactly one interior vertex $x$ is called a \emph{flower} with center $x$. \emph{The flower} at an interior vertex $x$ of a given 3-manifold is the flower with center $x$ which lies completely in the 3-manifold.
\end{defi}
As a consequence, in $Q(A_{N})$, in each flower every tetrahedron has exactly three adjacent 3-cells and every octahedron has exactly four adjacent 3-cells.\par
Examples for open 3-manifolds in $Q(A_{N})$ are the three-dimensional sub-lattices $Q(A_{3})$. Here, the flower at an interior vertex consists of eight tetrahedra (four black and four white ones) and six octahedra.\par
Examples of closed 3-manifolds in $Q(A_{N})$ are the set of facets of a 4-ambo-simplex (consisting of five tetrahedra) and the set of facets of a 4-ambo-simplex (consisting of five tetrahedra and five octahedra).\par
The elementary building blocks of 3-manifolds are so-called 4D~corners:
\begin{defi}[4D~corner]
A \emph{4D~corner} with center $x$ is a 3-manifold consisting of all facets of a 4-cell adjacent to $x$.
\end{defi}
In $Q(A_{N})$, there are two different types of 4D corners: a corner on a 4-simplex (consisting of a four tetrahedra) and a corner on a 4-ambo-simplex (consisting of two tetrahedra and three octahedra), see Appendix~\ref{sec:corners} for details.\par
The following combinatorial statement will be proven in Appendix~\ref{sec:proof}:
\begin{theo}\label{th:flower}
The flower at any interior vertex of any 3-manifold in $Q(A_{N})$ can be represented as a sum of 4D~corners in $Q(A_{N+2})$.
\end{theo}
Let $\Ell$ be a discrete 3-form on $Q(A_{N})$. The \emph{exterior derivative} $d\Ell$ is a discrete 4-form whose value at any 4-cell in $Q(A_{N})$ is the action functional of $\Ell$ on the 3-manifold consisting of the facets of the 4-cell. For our purposes, we consider discrete 3-forms $\Ell$ vanishing on all tetrahedra. In particular, we have
\[
d\Ell(\llfloor ijk\ell m\rrfloor)\equiv 0\quad\text{and}\quad d\Ell(\llceil ijk\ell m\rrceil)\equiv 0
\]
since a 4-simplices only contain tetrahedra. The exterior derivative on a black 4-ambo-simplex $\lfloor ijk\ell m\rfloor$ is given by
\begin{equation}\label{eq:extdev}
\begin{aligned}
\unbar{S}^{ijk\ell m}&:=d\Ell(\lfloor ijk\ell m\rfloor)\\
&\phantom{:}=\Ell([ijk\ell])+\Ell(-[ijkm])+\Ell([ij\ell m])+\Ell(-[ik\ell m])+\Ell([jk\ell m]).
\end{aligned}
\end{equation}
The exterior derivative on a white 4-ambo-simplex $\lceil ijk\ell m\rceil$ is given by
\begin{equation}\label{eq:extdev1}
\begin{aligned}
\bar{S}^{ijk\ell m}&:=d\Ell(\lceil ijk\ell m\rceil)\\
&\phantom{:}=\Ell(T_{m}[ijk\ell])+\Ell(-T_{\ell}[ijkm])+\Ell(T_{k}[ij\ell m])+\Ell(-T_{j}[ik\ell m])+\Ell(T_{i}[jk\ell m]).
\end{aligned}
\end{equation}\par
Accordingly, the Euler-Lagrange equations on black 4-ambo-simplices $\lfloor ijk\ell m\rfloor$ are
\begin{align}\label{eq:EL black}
\begin{alignedat}{5}
\frac{\partial\unbar{S}^{ijk\ell m}}{\partial x_{ij}}&=0,\quad&
\frac{\partial\unbar{S}^{ijk\ell m}}{\partial x_{ik}}&=0,\quad&
\frac{\partial\unbar{S}^{ijk\ell m}}{\partial x_{i\ell}}&=0,\quad&
\frac{\partial\unbar{S}^{ijk\ell m}}{\partial x_{im}}&=0,\quad&
\frac{\partial\unbar{S}^{ijk\ell m}}{\partial x_{jk}}&=0,\\
\frac{\partial\unbar{S}^{ijk\ell m}}{\partial x_{j\ell}}&=0,\quad&
\frac{\partial\unbar{S}^{ijk\ell m}}{\partial x_{jm}}&=0,\quad&
\frac{\partial\unbar{S}^{ijk\ell m}}{\partial x_{k\ell}}&=0,\quad&
\frac{\partial\unbar{S}^{ijk\ell m}}{\partial x_{km}}&=0,\quad&
\frac{\partial\unbar{S}^{ijk\ell m}}{\partial x_{\ell m}}&=0.
\end{alignedat}\\
\intertext{and the Euler-Lagrange equations on white 4-ambo-simplices $\lceil ijk\ell m\rceil$ are}\label{eq:EL white}
\begin{alignedat}{5}
\frac{\partial\bar{S}^{ijk\ell m}}{\partial x_{ijk}}&=0,\quad&
\frac{\partial\bar{S}^{ijk\ell m}}{\partial x_{ij\ell}}&=0,\quad&
\frac{\partial\bar{S}^{ijk\ell m}}{\partial x_{ijm}}&=0,\quad&
\frac{\partial\bar{S}^{ijk\ell m}}{\partial x_{ik\ell}}&=0,\quad&
\frac{\partial\bar{S}^{ijk\ell m}}{\partial x_{ikm}}&=0,\\
\frac{\partial\bar{S}^{ijk\ell m}}{\partial x_{i\ell m}}&=0,\quad&
\frac{\partial\bar{S}^{ijk\ell m}}{\partial x_{jk\ell}}&=0,\quad&
\frac{\partial\bar{S}^{ijk\ell m}}{\partial x_{jkm}}&=0,\quad&
\frac{\partial\bar{S}^{ijk\ell m}}{\partial x_{j\ell m}}&=0,\quad&
\frac{\partial\bar{S}^{ijk\ell m}}{\partial x_{k\ell m}}&=0.
\end{alignedat}
\end{align}
The last two systems are called \emph{corner equations}.\par
The following statement is an immediate consequence of Theorem~\ref{th:flower}:
\begin{theo}\label{cor:Corner}
For discrete every 3-form on $Q(A_{N})$ and every 3-manifold in $Q(A_{N})$ all corresponding Euler-Lagrange equations can be written as a sum of corner equations.
\end{theo}

\section{The dKP equation on \texorpdfstring{$Q(A_{N})$}{Q(A\_N)}}
We will now introduce the dKP equation on the root lattice $Q(A_{3})$. Every oriented octahedron~$[ijk\ell]$ ($i<j<k<\ell$) in $Q(A_{3})$ supports the equation
\begin{equation}\label{eq:dKP}
x_{ij}x_{k\ell}-x_{ik}x_{j\ell}+x_{i\ell}x_{jk}=0.
\end{equation}
We can extend this system in a consistent way (see~\cite{ABS3}) to the four-dimensional root lattice~$Q(A_{4})$ and higher-dimensional analogues, such that the five octahedral facets $[ijk\ell]$, $[jk\ell m]$, $-[ik\ell m]$, $[ijm\ell]$, and $-[ijkm]$ of the black 4-ambo-simplex $\lfloor ijk\ell m\rfloor$ support the equations
\begin{equation}\label{eq:dKPsystem black}
\begin{aligned}
&x_{ij}x_{k\ell}-x_{ik}x_{j\ell}+x_{i\ell}x_{jk}=0,\\
&x_{jk}x_{\ell m}-x_{j\ell}x_{km}+x_{jm}x_{k\ell}=0,\\
&x_{k\ell}x_{im}-x_{km}x_{i\ell}+x_{ik}x_{\ell m}=0,\\
&x_{\ell m}x_{ij}-x_{i\ell}x_{jm}+x_{j\ell}x_{im}=0,\\
&x_{im}x_{jk}-x_{jm}x_{ik}+x_{km}x_{ij}=0
\end{aligned}
\end{equation}
and the five octahedral facets $T_{m}[ijk\ell]$, $T_{i}[jk\ell m]$, $-T_{j}[ik\ell m]$, $T_{k}[ij\ell m]$, and $-T_{\ell}[ijkm]$ of the white 4-ambo-simplex $\lceil ijk\ell m\rceil$ support the equations
\begin{equation}\label{eq:dKPsystem white}
\begin{aligned}
&x_{ijm}x_{k\ell m}-x_{ikm}x_{j\ell m}+x_{i\ell m}x_{jkm}=0,\\
&x_{ijk}x_{i\ell m}-x_{ij\ell}x_{ikm}+x_{ijm}x_{ik\ell}=0,\\
&x_{jk\ell}x_{ijm}-x_{jkm}x_{ij\ell}+x_{ijk}x_{j\ell m}=0,\\
&x_{k\ell m}x_{ijk}-x_{ik\ell}x_{jkm}+x_{jk\ell}x_{ikm}=0,\\
&x_{i\ell m}x_{jk\ell}-x_{j\ell m}x_{ik\ell}+x_{k\ell m}x_{ij\ell}=0.
\end{aligned}
\end{equation}
In both systems one can derive one equation from another by cyclic permutations of indices~$(ijk\ell m)$.\par
We propose the following discrete 3-form $\Ell$ defined on oriented octahedra $[ijk\ell]$:
\begin{align}\label{eq:3-form}
&\begin{aligned}
&\Ell([ijk\ell]):=\frac{1}{2}\left(\Lambda\!\left(\frac{x_{ij}x_{k\ell}}{x_{ik}x_{j\ell}}\right)+\Lambda\!\left(\frac{x_{ik}x_{j\ell}}{x_{i\ell}x_{jk}}\right)+\Lambda\!\left(-\frac{x_{i\ell}x_{jk}}{x_{ij}x_{k\ell}}\right)\right),
\end{aligned}\\
\intertext{where}\label{eq:Lambda}
&\Lambda(z):=\lambda(z)-\lambda\!\left(\frac1z\right)\quad\text{and}\quad \lambda(z):=-\int_{0}^{z}\frac{\log|1-x|}{x}dx.
\end{align}
The discrete 3-form~\eqref{eq:3-form} has its motivation in~\cite{LNQ}. Indeed, in~\cite{LNQ}, the authors consider a similar discrete 3-form on the cubic lattice $\Z^{N}$. One can also consider our 3-form on the cubic lattice $\Z^{N}$. Then one would assign to each 3D~cube the 3-form at its inscribed octahedron. This 3-form differs from their one by an additive constant and a slightly different definition of the function $\lambda(z)$: they use the function
\begin{equation}\label{eq:dilog}
\Li_{2}(z):=-\int_{0}^{z}\frac{\log(1-x)}{x}dx
\end{equation}
instead of $\lambda(z)$. Our choice of $\lambda(z)$ allows us for a more precise consideration of the branches of the occurring logarithm.\par
Observe that the expression~\eqref{eq:3-form} only changes its sign under the cyclic permutation of indices~$(ijk\ell m)$. This follows from $\Lambda(z)=-\Lambda(z^{-1})$. As a consequence, the exterior derivatives $\unbar{S}^{ijk\ell m}$ and $\bar{S}^{ijk\ell m}$ defined in~\eqref{eq:extdev} and \eqref{eq:extdev1}, respectively, are invariant under the cyclic permutation of indices $(ijk\ell m)$. Therefore, one can obtain all corner equations in~\eqref{eq:EL black} and~\eqref{eq:EL white} by (iterated) cyclic permutation $(ijk\ell m)$ from
\[
\frac{\partial\unbar{S}^{ijk\ell m}}{\partial x_{ij}}=0,\quad\frac{\partial\unbar{S}^{ijk\ell m}}{\partial x_{ik}}=0,\quad\text{and}\quad\frac{\partial\bar{S}^{ijk\ell m}}{\partial x_{ijk}}=0,\quad\frac{\partial\bar{S}^{ijk\ell m}}{\partial x_{ij\ell}}=0.
\]\par
Let us study separately the corner equations on black and white 4-ambo-simplices. The corner equations which live on the black 4-ambo-simplex $\lfloor ijk\ell m\rfloor$ are given by
\begin{align}\notag
&\frac{\partial\unbar{S}^{ijk\ell m}}{\partial x_{ij}}=\frac{\partial\Ell([ijk\ell])}{\partial x_{ij}}+\frac{\partial\Ell(-[ijkm])}{\partial x_{ij}}+\frac{\partial\Ell([ij\ell m])}{\partial x_{ij}}=0\\
\intertext{and}\notag
&\frac{\partial\unbar{S}^{ijk\ell m}}{\partial x_{ik}}=\frac{\partial\Ell([ijk\ell])}{\partial x_{ik}}+\frac{\partial\Ell(-[ijkm])}{\partial x_{ik}}+\frac{\partial\Ell(-[ik\ell m])}{\partial x_{ik}}=0.
\end{align}
Explicitly, they read
\begin{align}\label{eq:corner black}
&\frac{1}{x_{ij}}\log|E_{ij}|=0\quad\text{and}\quad \frac{1}{x_{ik}}\log|E_{ik}|=0,\\
\intertext{where}\notag
&E_{ij}:=\frac{x_{ij}x_{k\ell}+x_{i\ell}x_{jk}}{x_{ij}x_{k\ell}-x_{ik}x_{j\ell}}\cdot\frac{x_{ij}x_{km}-x_{ik}x_{jm}}{x_{ij}x_{km}+x_{im}x_{jk}}\cdot\frac{x_{ij}x_{\ell m}+x_{im}x_{j\ell}}{x_{ij}x_{\ell m}-x_{i\ell}x_{jm}}\\
\intertext{and}\notag
&E_{ik}:=\frac{x_{ik}x_{j\ell}-x_{ij}x_{k\ell}}{x_{ik}x_{j\ell}-x_{i\ell}x_{jk}}\cdot\frac{x_{ik}x_{jm}-x_{im}x_{jk}}{x_{ik}x_{jm}-x_{ij}x_{km}}\cdot\frac{x_{ik}x_{\ell m}-x_{i\ell}x_{km}}{x_{ik}x_{\ell m}+x_{im}x_{k\ell}}.
\end{align}
For every corner equation~\eqref{eq:corner black} there are two classes of solutions, because any solution can either solve $E_{ij}=-1$ or $E_{ij}=1$.
Hereafter, we only consider solutions, where all fields $x_{ij}$ are non-zero (we call such solutions non-singular).\par
\begin{theo}\label{th:corner black}
Every solution of the system~\eqref{eq:EL black} solves either the system
\begin{align}\label{eq:DKP case}
&\begin{alignedat}{5}
E_{ij}&=-1,\quad&E_{ik}&=-1,\quad&E_{i\ell}&=-1,\quad&E_{im}&=-1,\quad&E_{jk}&=-1,\\
E_{j\ell}&=-1,\quad&E_{jm}&=-1,\quad&E_{k\ell}&=-1,\quad&E_{km}&=-1,\quad&E_{\ell m}&=-1
\end{alignedat}
\intertext{or the system}
&\begin{alignedat}{5}\label{eq:inverse DKP case}
E_{ij}&=1,\quad&E_{ik}&=1,\quad&E_{i\ell}&=1,\quad&E_{im}&=1,\quad&E_{jk}&=1,\\
E_{j\ell}&=1,\quad&E_{jm}&=1,\quad&E_{k\ell}&=1,\quad&E_{km}&=1,\quad&E_{\ell m}&=1.
\end{alignedat}
\end{align}
Furthermore, the system~\eqref{eq:DKP case} is equivalent to the system~\eqref{eq:dKPsystem black} (that is dKP on the corresponding black 4-ambo-simplex). The system~\eqref{eq:inverse DKP case} is equivalent to the system
\begin{equation}\label{eq:inverse dKPsystem black}
\begin{aligned}
&x_{ik}x_{i\ell}x_{jk}x_{j\ell}-x_{ij}x_{i\ell}x_{jk}x_{k\ell}+x_{ij}x_{ik}x_{j\ell}x_{k\ell}=0,\\
&x_{j\ell}x_{jm}x_{k\ell}x_{km}-x_{jk}x_{jm}x_{k\ell}x_{\ell m}+x_{jk}x_{j\ell}x_{km}x_{\ell m}=0,\\
&x_{km}x_{ik}x_{\ell m}x_{i\ell}-x_{k\ell}x_{ik}x_{\ell m}x_{im}+x_{k\ell}x_{km}x_{i\ell}x_{im}=0,\\
&x_{i\ell}x_{j\ell}x_{im}x_{jm}-x_{\ell m}x_{j\ell}x_{im}x_{ij}+x_{\ell m}x_{i\ell}x_{jm}x_{ij}=0,\\
&x_{jm}x_{km}x_{ij}x_{ik}-x_{im}x_{km}x_{ij}x_{jk}+x_{im}x_{jm}x_{ik}x_{jk}=0,
\end{aligned}
\end{equation}
which is the system~\eqref{eq:dKPsystem black} after the transformation $x\mapsto x^{-1}$ of fields (that is $\mathit{dKP}^{-}$ on the corresponding black 4-ambo-simplex).
\begin{proof}
Consider a solution $x$ of~\eqref{eq:EL black} that solves $E_{ij}=-1$ and $E_{jk}=-1$. We set
\begin{align}\label{eq:def aij}
&a_{ij}:=x_{\ell m}x_{ij}-x_{i\ell}x_{jm}+x_{j\ell}x_{im},\\\label{eq:def aik}
&a_{ik}:=x_{k\ell}x_{im}-x_{km}x_{i\ell}+x_{ik}x_{\ell m},\\
\intertext{and}\label{eq:def ajk}
&a_{jk}:=x_{jk}x_{\ell m}-x_{j\ell}x_{km}+x_{jm}x_{k\ell},
\end{align}
and use these equations to substitute $x_{ij}$, $x_{ik}$ and $x_{jk}$ in $E_{ij}=-1$ and $E_{jk}=-1$. Writing down the result in polynomial form, we get
\begin{align*}
&x_{\ell m}^{2}(a_{ij}+x_{i\ell}x_{jm}-x_{im}x_{j\ell})e_{ij}=0\\
\intertext{and}
&x_{\ell m}^{2}(a_{jk}+x_{j\ell}x_{km}-x_{jm}x_{k\ell})e_{jk}=0,
\end{align*}
where $e_{ij}$ and $e_{jk}$ are certain polynomials. Since for every solutions of~\eqref{eq:EL black} all fields are non-zero this leads us to $e_{ij}=0$ and $e_{jk}=0$. Computing the difference of the latter two equations we get
\begin{align*}
&a_{ij}x_{k\ell}x_{km}(a_{ij}+x_{i\ell}x_{jm}-x_{im}x_{j\ell})-a_{jk}x_{i\ell}x_{im}(a_{jk}+x_{j\ell}x_{km}-x_{jm}x_{k\ell})=0\\
\intertext{and, with the use of~\eqref{eq:def aij} and~\eqref{eq:def ajk},}
&x_{\ell m}(a_{ij}x_{ij}x_{k\ell}x_{km}-a_{jk}x_{jk}x_{i\ell}x_{im})=0,
\end{align*}
which depends on seven independent fields, i.e., no subset of six fields belong to one octahedron. Then comparing coefficients leads to $a_{ij}=a_{jk}=0$. Substituting
\begin{align*}
&x_{ij}=\frac{x_{i\ell}x_{jm}-x_{im}x_{j\ell}}{x_{\ell m}}\quad\text{and}\quad x_{jk}=\frac{x_{j\ell}x_{km}-x_{jm}x_{k\ell}}{x_{\ell m}}\\
\intertext{into $E_{ij}=-1$ and solving the resulting equation with respect to $x_{ik}$, we get}
&x_{ik}=\frac{x_{i\ell}x_{km}-x_{im}x_{k\ell}}{x_{\ell m}}.
\end{align*}
Substituting $x_{ij}$, $x_{ik}$ and $x_{jk}$ in $E_{ik}$ by using the last three equations, we get $E_{ik}=-1$.\par
Analogously, one can prove that, for a solution $x$ of~\eqref{eq:EL black} which solves $E_{ij}=-1$ and $E_{ik}=-1$, we have $E_{jk}=-1$, and for a solution $x$ of~\eqref{eq:EL black} which solves $E_{ik}=-1$ and $E_{i\ell}=-1$, we have $E_{k\ell}=-1$. Therefore, for every solution $x$ of~\eqref{eq:EL black} and for every white triangle $\{x_{\alpha},x_{\beta},x_{\gamma}\}$ on the black 4-ambo-simplex $\lfloor ijk\ell m\rfloor$ we proved the following: if $E_{\alpha}=-1$ and $E_{\beta}=-1$ then $E_{\gamma}=-1$, too.\par
On the other hand, one can easily see that $x$ solves $E_{ij}=1$ or $E_{jk}=1$ if and only if $x^{-1}$ solves $E_{ij}=-1$ or $E_{ik}=-1$, respectively. Therefore, we also know that, if $E_{\alpha}=1$ and $E_{\beta}=1$ then $E_{\gamma}=1$, too.\par
Summarizing, we proved that every solution $x$ of~\eqref{eq:EL black} solves either~\eqref{eq:DKP case} and then also~\eqref{eq:dKPsystem black} or~\eqref{eq:inverse DKP case} and then also~\eqref{eq:inverse dKPsystem black}.\par
Consider a non-singular solution $x$ of the system~\eqref{eq:dKPsystem black}. Then
\begin{align*}
&E_{ij}=\frac{x_{ij}x_{k\ell}+x_{i\ell}x_{jk}}{x_{ij}x_{k\ell}-x_{ik}x_{j\ell}}\cdot\frac{x_{ij}x_{km}-x_{ik}x_{jm}}{x_{ij}x_{km}+x_{im}x_{jk}}\cdot\frac{x_{ij}x_{\ell m}+x_{im}x_{j\ell}}{x_{ij}x_{\ell m}-x_{i\ell}x_{jm}}\\
&\phantom{E_{ij}}=\frac{x_{ik}x_{j\ell}}{-x_{i\ell}x_{jk}}\cdot\frac{-x_{im}x_{jk}}{x_{ik}x_{jm}}\cdot\frac{x_{i\ell}x_{jm}}{-x_{im}x_{j\ell}}=-1\\
\intertext{and}
&E_{ik}=\frac{x_{ik}x_{j\ell}-x_{ij}x_{k\ell}}{x_{ik}x_{j\ell}-x_{i\ell}x_{jk}}\cdot\frac{x_{ik}x_{jm}-x_{im}x_{jk}}{x_{ik}x_{jm}-x_{ij}x_{km}}\cdot\frac{x_{ik}x_{\ell m}-x_{i\ell}x_{km}}{x_{ik}x_{\ell m}+x_{im}x_{k\ell}}\\
&\phantom{E_{ik}}=\frac{x_{i\ell}x_{jk}}{x_{ij}x_{k\ell}}\cdot\frac{x_{ij}x_{km}}{x_{im}x_{jk}}\cdot\frac{x_{im}x_{k\ell}}{-x_{i\ell}x_{km}}=-1.
\end{align*}
This proves the equivalence of~\eqref{eq:DKP case} and~\eqref{eq:dKPsystem black} and also the equivalence of~\eqref{eq:inverse DKP case} and~\eqref{eq:inverse dKPsystem black} since $x$ solves $E_{ij}=-1$ or~\eqref{eq:dKPsystem black} if and only if $x^{-1}$ solves $E_{ij}=1$ or~\eqref{eq:inverse dKPsystem black}, respectively.
\end{proof}
\end{theo}
We will present the closure relation which can be seen as a criterion of integrability:
\begin{theo}[Closure relation]\label{th:closure black}
There holds:
\[
\unbar{S}^{ijk\ell m}\pm\frac{\pi^{2}}{4}=0
\]
on all solutions of~\eqref{eq:DKP case} and~\eqref{eq:inverse DKP case}, respectively. Therefore, one can redefine the 3-form $\Ell$ as
\[
\tilde{\Ell}([ijk\ell]):=\Ell([ijk\ell])\pm\frac{\pi^{2}}{4}
\]
in order to get $\unbar{S}^{ijk\ell m}=0$ on all solutions of~\eqref{eq:DKP case} and~\eqref{eq:inverse DKP case}, respectively.
\begin{proof}
The set of solutions $\mathcal{S}^{+}$ of~\eqref{eq:DKP case}, as well as the set of solutions $\Es^{-}$~\eqref{eq:inverse DKP case}, is a connected seven-dimensional algebraic manifold which can be parametrized by the set of variables $\{x_{ij},x_{ik},x_{i\ell},x_{im},x_{jk},x_{j\ell},x_{jm}\}$. We want to show that the directional derivatives of $\unbar{S}^{ijk\ell m}$ along tangent vectors of $\mathcal{S}^{\pm}$ vanish. It is easy to see that the stronger property $\grad \unbar{S}^{ijk\ell m}=0$ on $\mathcal{S}^{\pm}$, where we $\unbar{S}^{ijk\ell m}$ is considered as a function of ten variables $x_{ij}$, is a consequence of~\eqref{eq:DKP case}, respectively~\eqref{eq:inverse DKP case}. Therefore, the function $\unbar{S}^{ijk\ell m}$ is constant on $\mathcal{S}^{\pm}$.\par
To determine the value of $\unbar{S}^{ijk\ell m}$ on solutions of~\eqref{eq:DKP case}, we consider the constant solution of~\eqref{eq:dKPsystem black}
\begin{align}\label{eq:sol}
&\!\begin{aligned}
&x_{ij}=x_{jk}=x_{k\ell}=x_{\ell m}=x_{im}=a,\\
&x_{ik}=x_{j\ell}=x_{km}=x_{i\ell}=x_{jm}=-1,
\end{aligned}\\
\intertext{where}\notag
&a:=\frac12-\frac{\sqrt{5}}{2}.
\end{align}
(Indeed, for this point every equation from~\eqref{eq:dKPsystem black} looks like $a^{2}-1-a=0$.)
Therefore, this point satisfies~\eqref{eq:DKP case}, because \eqref{eq:dKPsystem black} and \eqref{eq:DKP case} are equivalent.\par
Consider the dilogarithm as defined in~\eqref{eq:dilog} and suppose that $z>1$. According to~\cite{lewin}, we derive:
\begin{align*}
&\Li_{2}(z)=-\Li_{2}(z^{-1})-\frac12\log^{2}z+\frac{\pi^{2}}{3}-i\pi\log z\\
\intertext{and}
&\Re\Li_{2}(z)=\Re\Li_{2}(ze^{i0})=-\frac{1}{2}\int_{0}^{z}\frac{\log(1-2x\cos 0+x^{2})}{x}dx=-\frac{1}{2}\int_{0}^{z}\frac{\log(1-x)^{2}}{x}dx\\
&\phantom{\Re\Li_{2}(z)}\, =-\int_{0}^{z}\frac{\log|1-x|}{x}dx=\lambda(z),
\end{align*}
where $\lambda(z)$ is the same function as in \eqref{eq:3-form}. Therefore, we have
\[
\lambda(z)=\begin{cases}
\Li_{2}(z),&z\leq1,\\
-\Li_{2}(z^{-1})-\dfrac12\log^{2}z+\dfrac{\pi^{2}}{3},&z>1.
\end{cases}
\]
By using the following special values~\cite{lewin}
\begin{alignat*}{2}
&\Li_{2}(a^{2})=\frac{\pi^{2}}{15}-\log^{2}(-a),\quad&
&\Li_{2}(-a)=\frac{\pi^{2}}{10}-\log^{2}(-a),\\
&\Li_{2}(a)=-\frac{\pi^{2}}{15}+\frac12\log^{2}(-a),\quad&
&\Li_{2}(a^{-1})=-\frac{\pi^{2}}{10}-\log^{2}(-a).
\end{alignat*}
a straightforward computation gives
\begin{align*}
&\Ell([ijk\ell])=\Ell(-[ijkm])=\Ell([ij\ell m])=\Ell(-[ik\ell m])=\Ell([jk\ell m])\\
&\phantom{\Ell([ijk\ell])}=\frac12(\Lambda(a^{2})+\Lambda(-a^{-1})+\Lambda(a^{-1}))=-\frac{\pi^{2}}{20}\\
\intertext{and}
&\unbar{S}^{ijk\ell m}=\Ell([ijk\ell])+\Ell(-[ijkm])+\Ell([ij\ell m])+\Ell(-[ik\ell m])+\Ell([jk\ell m])=-\frac{\pi^{2}}{4}.
\end{align*}
This is, because the expression for $\Ell([ijk\ell])$ (see~\eqref{eq:3-form}) changes the sign under the cyclic permutation of indices $(ijk\ell)$ and the solution is invariant under cyclic permutation of indices~$(ijk\ell m)$.\par
Let us now consider the second branch of solutions: one can easily see that
\begin{align}\label{eq:sol2}
&\!\begin{aligned}
&x_{ij}=x_{jk}=x_{k\ell}=x_{\ell m}=x_{im}=a^{-1},\\
&x_{ik}=x_{j\ell}=x_{km}=x_{i\ell}=x_{jm}=-1
\end{aligned}\\
\intertext{with}\notag
&a=\frac12-\frac{\sqrt{5}}{2}
\end{align}
is a solution of~\eqref{eq:inverse DKP case} and~\eqref{eq:inverse dKPsystem black}, because~\eqref{eq:sol} is a solution of~\eqref{eq:DKP case} and~\eqref{eq:dKPsystem black}. Therefore, on the solution~\eqref{eq:sol2} as well as on all other solutions of~\eqref{eq:inverse DKP case}, we have
\[
\unbar{S}^{ijk\ell m}=\frac{\pi^{2}}{4},
\]
where we used $\Lambda(z)=\lambda(z)-\lambda(z^{-1})$, and, therefore, $\Lambda(z^{-1})=-\Lambda(z)$.
\end{proof}
\end{theo}
Analogously, we get similar results for the white 4-ambo-simplex $\lceil ijk\ell m\rceil$. Here, the corner equations are:
\begin{align}\notag
&\frac{\partial\bar{S}^{ijk\ell m}}{\partial x_{ijk}}=\frac{\partial \Ell(T_{k}[ij\ell m])}{\partial x_{ijk}}+\frac{\partial \Ell(-T_{j}[ik\ell m])}{\partial x_{ijk}}+\frac{\partial \Ell(T_{i}[jk\ell m])}{\partial x_{ijk}}=0\\
\intertext{and}\notag
&\frac{\partial\bar{S}^{ijk\ell m}}{\partial x_{ij\ell}}=\frac{\partial \Ell(-T_{\ell}[ijkm])}{\partial x_{ij\ell}}+\frac{\partial \Ell(-T_{j}[ik\ell m])}{\partial x_{ij\ell}}+\frac{\partial \Ell(T_{i}[jk\ell m])}{\partial x_{ij\ell}}=0.\\
\intertext{Explicitly, they read}\label{eq:corner white}
&\frac{1}{x_{ijk}}\log|E_{ijk}|=0\quad\text{and}\quad \frac{1}{x_{ij\ell}}\log|E_{ij\ell}|=0,\\
\intertext{where}\notag
&E_{ijk}:=\frac{x_{ijk}x_{k\ell m}+x_{ikm}x_{jk\ell}}{x_{ijk}x_{k\ell m}-x_{ik\ell}x_{jkm}}\cdot\frac{x_{ijk}x_{j\ell m}-x_{ij\ell}x_{jkm}}{x_{ijk}x_{j\ell m}+x_{ijm}x_{jk\ell}}\cdot\frac{x_{ijk}x_{i\ell m}+x_{ijm}x_{ik\ell}}{x_{ijk}x_{i\ell m}-x_{ij\ell}x_{ikm}}\\
\intertext{and}\notag
&E_{ij\ell}:=\frac{x_{ij\ell}x_{k\ell m}-x_{ik\ell}x_{j\ell m}}{x_{ij\ell}x_{k\ell m}+x_{i\ell m}x_{jk\ell}}\cdot\frac{x_{ij\ell}x_{jkm}-x_{ijm}x_{jk\ell}}{x_{ij\ell}x_{jkm}-x_{ijk}x_{j\ell m}}\cdot\frac{x_{ij\ell}x_{ikm}-x_{ijk}x_{i\ell m}}{x_{ij\ell}x_{ikm}-x_{ijm}x_{ik\ell}}.
\end{align}
The analogue of Theorem~\ref{th:corner black} reads:
\begin{theo}\label{th:corner white}
Every solution of the system~\eqref{eq:EL white} solves either the system
\begin{align}\label{eq:DKP case white}
&\begin{alignedat}{5}
E_{ijk}&=-1,\quad&E_{ij\ell}&=-1,\quad&E_{ijm}&=-1,\quad&E_{ik\ell}&=-1,\quad&E_{ikm}&=-1,\\
E_{i\ell m}&=-1,\quad&E_{jk\ell}&=-1,\quad&E_{jkm}&=-1,\quad&E_{j\ell m}&=-1,\quad&E_{k\ell m}&=-1
\end{alignedat}
\intertext{or the system}
&\begin{alignedat}{5}\label{eq:inverse DKP case white}
E_{ijk}&=1,\quad&E_{ij\ell}&=1,\quad&E_{ijm}&=1,\quad&E_{ik\ell}&=1,\quad&E_{ikm}&=1,\\
E_{i\ell m}&=1,\quad&E_{jk\ell}&=1,\quad&E_{jkm}&=1,\quad&E_{j\ell m}&=1,\quad&E_{k\ell m}&=1.
\end{alignedat}
\end{align}
Furthermore the system~\eqref{eq:DKP case white} is equivalent to the system~\eqref{eq:dKPsystem white} (that is dKP on the corresponding white 4-ambo-simplex). The system~\eqref{eq:inverse DKP case white} is equivalent to the system
\begin{equation}\label{eq:inverse dKPsystem white}
\begin{aligned}
&x_{ikm}x_{i\ell m}x_{jkm}x_{j\ell m}-x_{ijm}x_{i\ell m}x_{jkm}x_{k\ell m}+x_{ijm}x_{ikm}x_{j\ell m}x_{k\ell m}=0,\\
&x_{ij\ell}x_{ijm}x_{ik\ell}x_{ikm}-x_{ijk}x_{ijm}x_{ik\ell}x_{i\ell m}+x_{ijk}x_{ij\ell}x_{ik\ell}x_{i\ell m}=0,\\
&x_{jkm}x_{ijk}x_{j\ell m}x_{ij\ell}-x_{jk\ell}x_{ijk}x_{j\ell m}x_{ijm}+x_{jk\ell}x_{jkm}x_{j\ell m}x_{ijm}=0,\\
&x_{ik\ell}x_{jk\ell}x_{ikm}x_{jkm}-x_{k\ell m}x_{jk\ell}x_{ikm}x_{ijk}+x_{k\ell m}x_{ik\ell}x_{ikm}x_{ijk}=0,\\
&x_{j\ell m}x_{k\ell m}x_{ij\ell}x_{ik\ell}-x_{i\ell m}x_{k\ell m}x_{ij\ell}x_{jk\ell}+x_{i\ell m}x_{j\ell m}x_{ij\ell}x_{jk\ell}=0,
\end{aligned}
\end{equation}
which is the system~\eqref{eq:dKPsystem white} after the transformation $x\mapsto x^{-1}$ of fields (that is $\mathit{dKP}^{-}$ on the corresponding white 4-ambo-simplex).
\end{theo}
The analogue of Theorem~\ref{th:closure black} reads:
\begin{theo}[Closure relation]\label{th:closure white}
There holds:
\[
\bar{S}^{ijk\ell m}\pm\frac{\pi^{2}}{4}=0
\]
on all solutions of~\eqref{eq:DKP case white} and~\eqref{eq:inverse DKP case white}, respectively. Therefore, one can redefine the 3-form $\Ell$ as
\[
\tilde{\Ell}([ijk\ell]):=\Ell([ijk\ell])\pm\frac{\pi^{2}}{4}
\]
in order to get $\bar{S}^{ijk\ell m}=0$ on all solutions of~\eqref{eq:DKP case white} and~\eqref{eq:inverse DKP case white}, respectively.
\end{theo}

\section{The cubic lattice \texorpdfstring{$\Z^{N}$}{Z\^{}N}}
We will now consider the relation between the elementary cells of the root lattice $Q(A_{N})$ and the cubic lattice $\Z^{N}$. The points of $Q(A_{N})$ and of $\Z^{N}$ are in a one-to-one correspondence via
\[
P_{i}:Q(A_{N})\to\Z^{N},\quad x(n_{0},\ldots,n_{i-1},n_{i},n_{i+1},\ldots,n_{N})\mapsto x(n_{0},\ldots,n_{i-1},n_{i+1},\ldots,n_{N}).
\]
In the present paper, we will always apply $P_{i}$ with $i<j,k,\ell,\ldots$\par
We denote by
\[
\{jk\ell\}:=\{x,x_{j},x_{k},x_{\ell},x_{jk},x_{j\ell},x_{k\ell},x_{jk\ell}\}
\]
the oriented 3D~cubes of $\Z^{N}$. We say that the 3D~cube $\{jk\ell\}$ is positively oriented if $j<k<\ell$. Any permutation of two indices changes the orientation to the opposite one. Also in this case, we always write the letters in the brackets in increasing order, so, e.g., in writing $\{jk\ell\}$ we assume that $j<k<\ell$ and avoid the notation $\{kj\ell\}$ or $\{j\ell k\}$ for the negatively oriented 3D~cube $-\{jk\ell\}$.\par
The object in $Q(A_{N})$ which corresponds to the 3D~cube $\{jk\ell\}$ is the sum of three adjacent 3-cells, namely
\begin{itemize}
\item the black tetrahedron $-T_{i}\lfloor ijk\ell\rfloor$ (see Figure~\ref{fig:3-cells}\subref{fig:black tetrahedron}),
\item the octahedron $[ijk\ell]$ (see Figure~\ref{fig:3-cells}\subref{fig:octahedron}),
\item and the white tetrahedron $-T_{\bar{\imath}}\lceil ijk\ell\rceil$ (see Figure~\ref{fig:3-cells}\subref{fig:white tetrahedron}).
\end{itemize}
It contains sixteen triangles and to every quadrilateral face of $\{jkl\}$ there corresponds a pair of these triangles containing one black and one white triangle. Here, the map $P_{i}$ reads as follows:
\[
x_{ii}\mapsto x,\quad x_{ij}\mapsto x_{j},\quad x_{jk}\mapsto x_{jk},\quad\text{and}\quad x_{\bar{\imath}jk\ell}\mapsto x_{jk\ell}.
\]\par
\begin{figure}[htb]
   \centering
   \subfloat[]{\label{fig:black tetrahedron}
   \begin{tikzpicture}[scale=0.85,inner sep=2]
      \node (x) at (0,0) [circle,fill,label=-135:$x_{ii}$] {};
      \node (x1) at (3,0) [circle,fill,label=-45:$x_{ij}$] {};
      \node (x2) at (1,1) [circle,fill,label=180:$x_{ik}$] {};
      \node (x3) at (0,3) [circle,fill,label=135:$x_{i\ell}$] {};
      \draw (x) to (x1);
      \draw [dashed] (x) to (x2);
      \draw (x) to (x3);
      \draw [dashed] (x1) to (x2);
      \draw (x1) to (x3);
      \draw [dashed] (x2) to (x3);
   \end{tikzpicture}
   }
   \subfloat[]{\label{fig:octahedron}
   \begin{tikzpicture}[scale=0.85,inner sep=2]
      \node (x1) at (3,0) [circle,fill,label=-45:$x_{ij}$] {};
      \node (x2) at (1,1) [circle,fill,label=-135:$x_{ik}$] {};
      \node (x3) at (0,3) [circle,fill,label=135:$x_{i\ell}$] {};
      \node (x12) at (4,1) [circle,fill,label=-45:$x_{jk}$] {};
      \node (x13) at (3,3) [circle,fill,label=45:$x_{j\ell}$] {};
      \node (x23) at (1,4) [circle,fill,label=135:$x_{k\ell}$] {};
      \draw (x1) to (x2);
      \draw (x1) to (x3);
      \draw (x1) to (x12);
      \draw (x1) to (x13);
      \draw (x2) to (x3);
      \draw [dashed] (x2) to (x12);
      \draw [dashed] (x2) to (x23);
      \draw (x3) to (x13);
      \draw (x3) to (x23);
      \draw (x12) to (x13);
      \draw [dashed] (x12) to (x23);
      \draw (x13) to (x23);
   \end{tikzpicture}
   }
   \subfloat[]{\label{fig:white tetrahedron}
      \begin{tikzpicture}[scale=0.85,inner sep=2]
      \node (x12) at (4,1) [circle,fill,label=-45:$x_{jk}$] {};
      \node (x13) at (3,3) [circle,fill,label=0:$x_{j\ell}$] {};
      \node (x23) at (1,4) [circle,fill,label=135:$x_{k\ell}$] {};
      \node (x123) at (4,4) [circle,fill,label=45:$x_{\bar{\imath}jk\ell}$] {};
      \draw (x12) to (x13);
      \draw (x12) to (x123);
      \draw (x12) to (x23);
      \draw (x13) to (x23);
      \draw (x13) to (x123);
      \draw (x23) to (x123);
   \end{tikzpicture}
   }\\
   \subfloat[]{\label{fig:cube}
   \begin{tikzpicture}[scale=0.85,inner sep=2]
      \node (x) at (0,0) [circle,fill,label=-135:$x_{ii}$] {};
      \node (x1) at (3,0) [circle,fill,label=-45:$x_{ij}$] {};
      \node (x2) at (1,1) [circle,fill,label=180:$x_{ik}$] {};
      \node (x3) at (0,3) [circle,fill,label=135:$x_{i\ell}$] {};
      \node (x12) at (4,1) [circle,fill,label=-45:$x_{jk}$] {};
      \node (x13) at (3,3) [circle,fill,label=0:$x_{j\ell}$] {};
      \node (x23) at (1,4) [circle,fill,label=135:$x_{k\ell}$] {};
      \node (x123) at (4,4) [circle,fill,label=45:$x_{\bar{\imath}jk\ell}$] {};
      \draw (x) to (x1);
      \draw [dashed] (x) to (x2);
      \draw (x) to (x3);
      \draw [dashed] (x1) to (x2);
      \draw (x1) to (x3);
      \draw (x1) to (x12);
      \draw (x1) to (x13);
      \draw [dashed] (x2) to (x3);
      \draw [dashed] (x2) to (x12);
      \draw [dashed] (x2) to (x23);
      \draw (x3) to (x13);
      \draw (x3) to (x23);
      \draw (x12) to (x13);
      \draw (x12) to (x123);
      \draw [dashed] (x12) to (x23);
      \draw (x13) to (x23);
      \draw (x13) to (x123);
      \draw (x23) to (x123);
   \end{tikzpicture}
   }\qquad
   \subfloat[]{\label{fig:cube1}
   \begin{tikzpicture}[scale=0.85,inner sep=2]
      \node (x) at (0,0) [circle,fill,label=-135:$x$] {};
      \node (x1) at (3,0) [circle,fill,label=-45:$x_{j}$] {};
      \node (x2) at (1,1) [circle,fill,label=180:$x_{k}$] {};
      \node (x3) at (0,3) [circle,fill,label=135:$x_{\ell}$] {};
      \node (x12) at (4,1) [circle,fill,label=-45:$x_{jk}$] {};
      \node (x13) at (3,3) [circle,fill,label=0:$x_{j\ell}$] {};
      \node (x23) at (1,4) [circle,fill,label=135:$x_{k\ell}$] {};
      \node (x123) at (4,4) [circle,fill,label=45:$x_{jk\ell}$] {};
      \draw (x) to (x1);
      \draw [dashed] (x) to (x2);
      \draw (x) to (x3);
      \draw (x1) to (x12);
      \draw (x1) to (x13);
      \draw [dashed] (x2) to (x12);
      \draw [dashed] (x2) to (x23);
      \draw (x3) to (x13);
      \draw (x3) to (x23);
      \draw (x12) to (x123);
      \draw (x13) to (x123);
      \draw (x23) to (x123);
   \end{tikzpicture}
   }
   \caption{Three adjacent 3-cells of the lattice $Q(A_{N})$: \protect\subref{fig:black tetrahedron} black tetrahedron $-T_{i}\lfloor ijk\ell\rfloor$, \protect\subref{fig:octahedron} octahedron $[ijk\ell]$, \protect\subref{fig:white tetrahedron} white tetrahedron $-T_{\bar{\imath}}\lceil ijk\ell\rceil$. The sum \protect\subref{fig:cube} of these 3-cells corresponds to a 3D~cube~\protect\subref{fig:cube1}.}
   \label{fig:3-cells}
\end{figure}
As a four-dimensional elementary cell of $\Z^{N}$, we consider an oriented 4D~cube
\[
\{jk\ell m\}:=\{x,x_{j},x_{k},x_{\ell},x_{m},x_{jk},x_{j\ell},x_{jm},x_{k\ell},x_{km},x_{\ell m},x_{jk\ell},x_{jkm},x_{j\ell m},x_{k\ell m},x_{jk\ell m}\}.
\]
The 4D~cube $\{jk\ell m\}$ corresponds to the sum of four 4-cells in $Q(A_{N})$:
\begin{itemize}
\item the black 4-simplex $-T_{i}\llfloor ijk\ell m\rrfloor$,
\item the black 4-ambo-simplex $\lfloor ijk\ell m\rfloor$,
\item the white 4-ambo-simplex $-T_{\bar{\imath}}\lceil ijk\ell m\rceil$, and
\item the white 4-simplex $T_{\bar{\imath}}T_{\bar{\imath}}\llceil ijk\ell m\rrceil$
\end{itemize}
(see Figure~\ref{fig:4D cube}). It contains sixteen tetrahedra (eight black and eight white ones) and eight octahedra. Here, the map $P_{i}$ reads as follows:
\[
x_{ii}\mapsto x,\quad x_{ij}\mapsto x_{j},\quad x_{jk}\mapsto x_{jk},\quad x_{\bar{\imath}jk\ell}\mapsto x_{jk\ell},\quad\text{and}\quad x_{\bar{\imath}\bar{\imath}jk\ell m}\mapsto x_{jk\ell m}.
\]\par
\begin{figure}[htbp]
   \centering
   \begin{tikzpicture}[inner sep=2]
      \node (x) at (0,0) [circle,fill,label=135:$x_{ii}$] {};
      \node (x1) at (2,0) [circle,fill,label=-135:$x_{ij}$] {};
      \node (x2) at (1,0.5) [circle,fill,label=45:$x_{ik}$] {};
      \node (x3) at (0,2) [circle,fill,label=-45:$x_{i\ell}$] {};
      \node (x4) at (-3.25,-2.5) [circle,fill,label=-135:$x_{im}$] {};
      \node (x12) at (3,0.5) [circle,fill,label=45:$x_{jk}$] {};
      \node (x13) at (2,2) [circle,fill,label=-135:$x_{j\ell}$] {};
      \node (x14) at (2.75,-2.5) [circle,fill,label=-45:$x_{jm}$] {};
      \node (x23) at (1,2.5) [circle,fill,label=45:$x_{k\ell}$] {};
      \node (x24) at (-0.25,-0.75) [circle,fill,label=-45:$x_{km}$] {};
      \node (x34) at (-3.25,3.5) [circle,fill,label=135:$x_{\ell m}$] {};
      \node (x123) at (3,2.5) [circle,fill,label=-45:$x_{\bar{\imath}jk\ell}$] {};
      \node (x124) at (5.75,-0.75) [circle,fill,label=-45:$x_{\bar{\imath}jkm}$] {};
      \node (x134) at (2.75,3.5) [circle,fill,label=135:$x_{\bar{\imath}j\ell m}$] {};
      \node (x234) at (-0.25,5) [circle,fill,label=135:$x_{\bar{\imath}k\ell m}$] {};
      \node (x1234) at (5.75,5) [circle,fill,label=45:$x_{\bar{\imath}\bar{\imath}jk\ell m}$] {};
      \draw (x) to (x1) to (x12) to (x123) to (x23) to (x3) to (x);
      \draw (x4) to (x14) to (x124) to (x1234) to (x234) to (x34) to (x4);
      \draw (x1) to (x13) to (x3);
      \draw (x14) to (x134) to (x34);
      \draw (x13) to (x123);
      \draw (x134) to (x1234);
      \draw [dashed] (x) to (x2) to (x12);
      \draw [dashed] (x4) to (x24) to (x124);
      \draw [dashed] (x2) to (x23);
      \draw [dashed] (x24) to (x234);
      \draw [dotted,thick] (x) to (x4);
      \draw [dotted,thick] (x1) to (x14);
      \draw [dotted,thick] (x2) to (x24);
      \draw [dotted,thick] (x3) to (x34);
      \draw [dotted,thick] (x12) to (x124);
      \draw [dotted,thick] (x13) to (x134);
      \draw [dotted,thick] (x23) to (x234);
      \draw [dotted,thick] (x123) to (x1234);
      \node (x) at (0,0) [circle,fill,label={135,fill=white:$x_{ii}$}] {};
   \end{tikzpicture}
   \caption{The sum of the black 4-simplex $-T_{i}\llfloor ijk\ell m\rrfloor$, the adjacent black 4-ambo-simplex $\lfloor ijk\ell m\rfloor$, the adjacent white 4-ambo-simplex $-T_{\bar{\imath}}\lceil ijk\ell m\rceil$, and the adjacent white 4-simplex $T_{\bar{\imath}}T_{\bar{\imath}}\llceil ijk\ell m\rrceil$ corresponds to the 4D~cube $\{jk\ell m\}$.}
   \label{fig:4D cube}
\end{figure}
Also in the cubic case there is an easy recipe to obtain the orientation of the facets of an (oriented) 4D~cube: on every index between the brackets we put alternately a ``$+$'' and a ``$-$'' starting with a ``$+$'' on the last index. Then we get each facet by deleting one index and putting the corresponding sign in front of the bracket. For instance., the 4D~cube
{
\setlength\arraycolsep{0pt}
\renewcommand{\arraystretch}{0.5}
\begin{equation*}
\begin{array}{cccccc}
&-&+&-&+&\\
\{&j&k&\ell&m&\}
\end{array}
\end{equation*}
has the eight 3D~facets: $\{jk\ell\}$, $-\{jkm\}$, $\{j\ell m\}$, $-\{k\ell m\}$ and the opposite ones $-T_{m}\{jk\ell\}$, $T_{\ell}\{jkm\}$, $-T_{k}\{j\ell m\}$, and $T_{j}\{k\ell m\}$.\par}
As a consequence of Definition~\ref{def:flower}, in each flower in $\Z^{N}$, every 3D~cube has exactly four adjacent 3D~cubes.\par
We will now prove the analogue of Theorem~\ref{cor:Corner}. This proof is easier than the one for $Q(A_{N})$, because of the simpler combinatorial structure.
\begin{theo}\label{th:flower cube}
The flower at any interior vertex of any 3-manifold in $\Z^{N}$ can be represented as a sum of 4D~corners in $\Z^{N+1}$.
\begin{proof}
Set $M:=N+1$ and consider the flower of an interior vertex $x$ of an arbitrary 3-manifold in $\Z^{N}$. Over each 3D~corner $\{jk\ell\}$ (petal) of the flower, we can build a 4D~corner adjacent to $x$ on the 4D~cube $\{jk\ell M\}$. Then the ‘vertical’ 3D~cubes coming from two successive petals of the flower carry opposite orientations, so that all ‘vertical’ squares cancel away from the sum of the 4D~corners.
\end{proof}
\end{theo}
Let $\ELL$ be a discrete 3-form on $\Z^{N}$. The \emph{exterior derivative} $d\ELL$ is a discrete 4-form whose value at any 4D~cube in $\Z^{N}$ is the action functional of $\ELL$ on the 3-manifold consisting of the facets of the 4D~cube:
\begin{multline*}
S^{jk\ell m}:=d\ELL(\{jk\ell m\})=\ELL(\{jk\ell\})+\ELL(-\{jkm\})+\ELL(\{j\ell m\})+\ELL(-\{k\ell m\})\\
+\ELL(-T_{m}\{jk\ell\})+\ELL(T_{\ell}\{jkm\})+\ELL(-T_{k}\{j\ell m\})+\ELL(T_{j} \{k\ell m\}).
\end{multline*}
Accordingly, the Euler-Lagrange equations on the 4D~cube $\{jk\ell m\}$ are given by
\begin{equation}\label{eq:EL cube}
\begin{alignedat}{6}
\frac{\partial S^{jk\ell m}}{\partial x}&=0,\\
\frac{\partial S^{jk\ell m}}{\partial x_{j}}&=0,\quad&
\frac{\partial S^{jk\ell m}}{\partial x_{k}}&=0,\quad&
\frac{\partial S^{jk\ell m}}{\partial x_{\ell}}&=0,\quad&
\frac{\partial S^{jk\ell m}}{\partial x_{m}}&=0,\\
\frac{\partial S^{jk\ell m}}{\partial x_{jk}}&=0,\quad&
\frac{\partial S^{jk\ell m}}{\partial x_{j\ell}}&=0,\quad&
\frac{\partial S^{jk\ell m}}{\partial x_{jm}}&=0,\quad&
\frac{\partial S^{jk\ell m}}{\partial x_{k\ell}}&=0,\quad&
\frac{\partial S^{jk\ell m}}{\partial x_{km}}&=0,\quad&
\frac{\partial S^{jk\ell m}}{\partial x_{\ell m}}&=0,\\
\frac{\partial S^{jk\ell m}}{\partial x_{jk\ell}}&=0,\quad&
\frac{\partial S^{jk\ell m}}{\partial x_{jkm}}&=0,\quad&
\frac{\partial S^{jk\ell m}}{\partial x_{j\ell m}}&=0,\quad&
\frac{\partial S^{jk\ell m}}{\partial x_{k\ell m}}&=0,\\
\frac{\partial S^{jk\ell m}}{\partial x_{jk\ell m}}&=0.
\end{alignedat}
\end{equation}
They are called \emph{corner equations}.\par
The following statement is an immediate consequence of Theorem~\ref{th:flower cube}:
\begin{theo}\label{cor:Corner cubic}
For every discrete 3-form on $\Z^{N}$ and every 3-manifold in $\Z^{N}$ all corresponding Euler-Lagrange equations can be written as a sum of corner equations.
\end{theo}

\section{The dKP equation on \texorpdfstring{$\Z^{N}$}{Z\^{}N}}
On the 3D~cube $\{jk\ell\}$ in $\Z^{3}$ ($j<k<\ell$) we put the equation
\begin{equation}\label{eq:dKP cube}
x_{j}x_{k\ell}-x_{k}x_{j\ell}+x_{\ell}x_{jk}=0.
\end{equation}
We can extend this system in a consistent way (see~\cite{ABS3}) to the four-dimensional cubic lattice $\Z^{4}$ and its higher-dimensional analogues, such that the eight facets $\{jk\ell\}$, $-\{jkm\}$, $\{j\ell m\}$, $-\{k\ell m\}$, $-T_{m}\{jk\ell\}$, $T_{\ell}\{jkm\}$, $-T_{k}\{j\ell m\}$, $T_{j} \{k\ell m\}$ of a 4D~cube $\{jk\ell m\}$ carry the equations
\begin{equation}\label{eq:dKPsystem cube}
\begin{alignedat}{2}
&x_{j}x_{k\ell}-x_{k}x_{j\ell}+x_{\ell}x_{jk}=0,&\qquad&x_{jm}x_{k\ell m}-x_{km}x_{j\ell m}+x_{\ell m}x_{jkm}=0,\\
&x_{j}x_{km}-x_{k}x_{jm}+x_{m}x_{jk}=0,&\qquad&x_{jk}x_{k\ell m}-x_{k\ell}x_{jkm}+x_{km}x_{jk\ell}=0,\\
&x_{j}x_{\ell m}-x_{\ell}x_{jm}+x_{m}x_{j\ell}=0,&\qquad&x_{j\ell}x_{k\ell m}-x_{k\ell}x_{j\ell m}+x_{\ell m}x_{jk\ell}=0,\\
&x_{k}x_{\ell m}-x_{\ell}x_{km}+x_{m}x_{k\ell}=0,&\qquad&x_{jk}x_{j\ell m}-x_{j\ell}x_{jkm}+x_{jm}x_{jk\ell}=0.
\end{alignedat}
\end{equation}
Note that, in the four equations in the left column, the fields with one index always appear with increasing order of indices. The equations in the right column are shifted copies of the ones in the left column. One can derive the system~\eqref{eq:dKPsystem cube} from the system of dKP equations~\eqref{eq:dKPsystem black} on the black 4-ambo-simplex $\lfloor ijk\ell m\rfloor$ and the system of dKP equations~\eqref{eq:dKPsystem white} on the white 4-ambo-simplex $T_{\bar{\imath}}\lceil ijkm\ell\rceil$, by removing the equations on the octahedra $[jk\ell m]$ and $[jkm\ell]$, respectively, from both systems and applying the transformation $P_{i}$ to the fields in the remaining eight equations.\par
We propose the discrete 3-form $\ELL$ defined as
\[
\ELL(\{jk\ell\}):=\Ell(P_{i}[ijk\ell]),
\]
where $\Ell$ is the discrete 3-form on the root lattice $Q(A_{N})$ (see~\eqref{eq:3-form}).\par
For this discrete 3-form, there are no corner equations on the 4D~cube $\{jk\ell m\}$ centered at $x$ and $x_{jk\ell m}$ since $S^{jk\ell m}$ does not depend on these two variables. The remaining corner equations from~\eqref{eq:EL cube} are given by
\begin{equation}\label{eq:corner cube}
\begin{alignedat}{2}
\frac{\partial S^{jk\ell m}}{\partial x_{j}}&=\frac{\partial\ELL(\{jk\ell\})}{\partial x_{j}}+\frac{\ELL(-\{jkm\})}{\partial x_{j}}+\frac{\partial\ELL(\{j\ell m\})}{\partial x_{j}}+\underbrace{\frac{\partial\ELL(T_{j} \{k\ell m\})}{\partial x_{j}}}_{\equiv0}\\
&=\frac{\partial\Ell(P_{i}[ijk\ell])}{\partial x_{j}}+\frac{\partial\Ell(-P_{i}[ijkm])}{\partial x_{j}}+\frac{\partial\Ell(P_{i}[ij\ell m])}{\partial x_{j}}=\frac{1}{x_{j}}\log|\E_{j}|=0,\\
\frac{\partial S^{jk\ell m}}{\partial x_{jk}}&=\frac{\partial\ELL(\{jk\ell\})}{\partial x_{jk}}+\frac{\partial\ELL(-\{jkm\})}{\partial x_{jk}}+\frac{\partial\ELL(-T_{k}\{j\ell m\})}{\partial x_{jk}}+\frac{\partial\ELL(T_{j}\{k\ell m\})}{\partial x_{jk}}\\
&=\frac{\partial\Ell(P_{i}[ijk\ell])}{\partial x_{jk}}+\frac{\partial\Ell(-P_{i}[ijkm])}{\partial x_{jk}}+\frac{\partial\Ell(-P_{i}T_{\bar{\imath}}T_{k}[ij\ell m])}{\partial x_{jk}}+\frac{\partial\Ell(P_{i}T_{\bar{\imath}}T_{j}[ik\ell m])}{\partial x_{jk}}\\
&=\frac{1}{x_{jk}}\log\left|\frac{\unbar{\E}_{jk}}{\bar{\E}_{jk}}\right|=0,\\
\frac{\partial S^{jk\ell m}}{\partial x_{jk\ell}}&=\underbrace{\frac{\partial\ELL(\{jk\ell\})}{\partial x_{jk\ell}}}_{\equiv0}+\frac{\partial\ELL(T_{\ell}\{jkm\})}{\partial x_{jk\ell}}+\frac{\partial\ELL(-T_{k}\{j\ell m\})}{\partial x_{jk\ell}}+\frac{\partial\ELL(T_{j} \{k\ell m\})}{\partial x_{jk\ell}}\\
&=\frac{\partial\Ell(P_{i}T_{\bar{\imath}}T_{\ell}[ijkm])}{\partial x_{jk\ell}}+\frac{\partial\Ell(-P_{i}T_{\bar{\imath}}T_{k}[ij\ell m])}{\partial x_{jk\ell}}+\frac{\partial\Ell(P_{i}T_{\bar{\imath}}T_{j}[ik\ell m])}{\partial x_{jk\ell}}\\
&=\frac{1}{x_{jk\ell}}\log\left|\frac{1}{\E_{jk\ell}}\right|=0,\\
\end{alignedat}
\end{equation}
where $\E_{j}$ and $\unbar{\E}_{jk}$ are obtained from $E_{ij}$ and $E_{jk}$, respectively, by using the transformation $P_{i}$ of fields, and $\bar{\E}_{jk}$ and $\E_{jk\ell}$ are obtained from $E_{ijk}$ and $E_{jk\ell}$, respectively, by using the transformation $P_{i}\circ T_{\bar{\imath}}$ of fields.\par
Hereafter, we only consider solutions, where all fields are non-zero (we call these solutions non-singular). As in the case of the root lattice~$Q(A_{N})$ every corner equation has two classes of solutions.
\begin{theo}
Every solution of the system~\eqref{eq:EL cube} solves either the system
\begin{align}\label{eq:DKP case cube}
&\begin{alignedat}{6}
\E_{j}&=-1,\quad&\E_{k}&=-1,\quad&\E_{\ell}&=-1,\quad&\E_{m}&=-1,\\
\unbar{\E}_{jk}&=-1,\quad&\unbar{\E}_{j\ell}&=-1,\quad&\unbar{\E}_{jm}&=-1,\quad&\unbar{\E}_{k\ell}&=-1,\quad&\unbar{\E}_{km}&=-1,\quad&\unbar{\E}_{\ell m}&=-1,\\
\bar{\E}_{jk}&=-1,\quad&\bar{\E}_{j\ell}&=-1,\quad&\bar{\E}_{jm}&=-1,\quad&\bar{\E}_{k\ell}&=-1,\quad&\bar{\E}_{km}&=-1,\quad&\bar{\E}_{\ell m}&=-1,\\
\E_{jk\ell}&=-1,\quad&\E_{jkm}&=-1,\quad&\E_{j\ell m}&=-1,\quad&\E_{k\ell m}&=-1
\end{alignedat}
\intertext{or the system}
&\begin{alignedat}{6}\label{eq:inverse DKP case cube}
\E_{j}&=1,\quad&\E_{k}&=1,\quad&\E_{\ell}&=1,\quad&\E_{m}&=1,\\
\unbar{\E}_{jk}&=1,\quad&\unbar{\E}_{j\ell}&=1,\quad&\unbar{\E}_{jm}&=1,\quad&\unbar{\E}_{k\ell}&=1,\quad&\unbar{\E}_{km}&=1,\quad&\unbar{\E}_{\ell m}&=1,\\
\bar{\E}_{jk}&=1,\quad&\bar{\E}_{j\ell}&=1,\quad&\bar{\E}_{jm}&=1,\quad&\bar{\E}_{k\ell}&=1,\quad&\bar{\E}_{km}&=1,\quad&\bar{\E}_{\ell m}&=1,\\
\E_{jk\ell}&=1,\quad&\E_{jkm}&=1,\quad&\E_{j\ell m}&=1,\quad&\E_{k\ell m}&=1.
\end{alignedat}
\end{align}
Furthermore the system~\eqref{eq:DKP case cube} is equivalent to the system~\eqref{eq:dKPsystem cube} (this is dKP on the corresponding 4D~cube). The system~\eqref{eq:inverse DKP case cube} is equivalent to the system
\begin{equation}\label{eq:inverse dKPsystem cube}
\begin{aligned}
&x_{k}x_{\ell}x_{jk}x_{j\ell}-x_{j}x_{\ell}x_{jk}x_{k\ell}+x_{j}x_{k}x_{j\ell}x_{k\ell}=0,\\
&x_{k}x_{m}x_{jk}x_{jm}-x_{j}x_{m}x_{jk}x_{km}+x_{j}x_{k}x_{jm}x_{km}=0,\\
&x_{\ell}x_{m}x_{j\ell}x_{jm}-x_{j}x_{m}x_{j\ell}x_{\ell m}+x_{j}x_{\ell}x_{jm}x_{\ell m}=0,\\
&x_{\ell}x_{m}x_{k\ell}x_{km}-x_{k}x_{m}x_{k\ell}x_{\ell m}+x_{k}x_{\ell}x_{km}x_{\ell m}=0,\\
&x_{km}x_{\ell m}x_{jkm}x_{j\ell m}-x_{jm}x_{\ell m}x_{jkm}x_{k\ell m}+x_{jm}x_{km}x_{j\ell m}x_{k\ell m}=0,\\
&x_{k\ell}x_{\ell m}x_{jk\ell}x_{j\ell m}-x_{j\ell}x_{\ell m}x_{jk\ell}x_{k\ell m}+x_{j\ell}x_{k\ell}x_{j\ell m}x_{k\ell m}=0,\\
&x_{k\ell}x_{km}x_{jk\ell}x_{jkm}-x_{jk}x_{km}x_{jk\ell}x_{k\ell m}+x_{jk}x_{k\ell}x_{jkm}x_{k\ell m}=0,\\
&x_{j\ell}x_{jm}x_{jk\ell}x_{jkm}-x_{jk}x_{jm}x_{jk\ell}x_{j\ell m}+x_{jk}x_{j\ell}x_{jkm}x_{j\ell m}=0,
\end{aligned}
\end{equation}
which is the system~\eqref{eq:dKPsystem cube} after the transformation $x\mapsto x^{-1}$ of fields (this is $\mathit{dKP}^{-}$ on the corresponding 4D~cube).
\begin{proof}
Let $x$ be a solution of the system~\eqref{eq:EL cube} such that $\E_{j}=-1$ and $\E_{k}=-1$. Then we know from the proof of Theorem~\ref{th:corner black} that
\begin{align*}
&\begin{alignedat}{6}
\E_{j}&=-1,\quad&\E_{k}&=-1,\quad&\E_{\ell}&=-1,\quad&\E_{m}&=-1,\\
\unbar{\E}_{jk}&=-1,\quad&\unbar{\E}_{j\ell}&=-1,\quad&\unbar{\E}_{jm}&=-1,\quad&\unbar{\E}_{k\ell}&=-1,\quad&\unbar{\E}_{km}&=-1,\quad&\unbar{\E}_{\ell m}&=-1
\end{alignedat}
\intertext{and that the latter system is equivalent to}
&x_{j}x_{k\ell}-x_{k}x_{j\ell}+x_{\ell}x_{jk}=0,\\
&x_{j}x_{km}-x_{k}x_{jm}+x_{m}x_{jk}=0,\\
&x_{j}x_{\ell m}-x_{\ell}x_{jm}+x_{m}x_{j\ell}=0,\\
&x_{k}x_{\ell m}-x_{\ell}x_{km}+x_{m}x_{k\ell}=0,\\
&x_{jk}x_{\ell m}-x_{j\ell}x_{km}+x_{jm}x_{k\ell}=0.
\end{align*}
On the other hand, if we consider a solution $x$ of~\eqref{eq:EL cube} such that $\E_{j}=1$ and $\E_{k}=1$, we know from the proof of Theorem~\ref{th:corner black} that
\begin{align*}
&\begin{alignedat}{6}
\E_{j}&=1,\quad&\E_{k}&=1,\quad&\E_{\ell}&=1,\quad&\E_{m}&=1,\\
\unbar{\E}_{jk}&=1,\quad&\unbar{\E}_{j\ell}&=1,\quad&\unbar{\E}_{jm}&=1,\quad&\unbar{\E}_{k\ell}&=1,\quad&\unbar{\E}_{km}&=1,\quad&\unbar{\E}_{\ell m}&=1
\end{alignedat}
\intertext{and that the latter system is equivalent to}
&x_{k}x_{\ell}x_{jk}x_{j\ell}-x_{j}x_{\ell}x_{jk}x_{k\ell}+x_{j}x_{k}x_{j\ell}x_{k\ell}=0,\\
&x_{k}x_{m}x_{jk}x_{jm}-x_{j}x_{m}x_{jk}x_{km}+x_{j}x_{k}x_{jm}x_{km}=0,\\
&x_{\ell}x_{m}x_{j\ell}x_{jm}-x_{j}x_{m}x_{j\ell}x_{\ell m}+x_{j}x_{\ell}x_{jm}x_{\ell m}=0,\\
&x_{\ell}x_{m}x_{k\ell}x_{km}-x_{k}x_{m}x_{k\ell}x_{\ell m}+x_{k}x_{\ell}x_{km}x_{\ell m}=0,\\
&x_{j\ell}x_{jm}x_{k\ell}x_{km}-x_{jk}x_{jm}x_{k\ell}x_{\ell m}+x_{jk}x_{j\ell}x_{km}x_{\ell m}=0.
\end{align*}
Now, let $x$ be a solution of the system~\eqref{eq:EL cube} such that $\E_{jk\ell}=-1$ and $\E_{jkm}=-1$. Then we know from the proof of Theorem~\ref{th:corner white} that
\begin{align*}
&\begin{alignedat}{6}
\bar{\E}_{jk}&=1,\quad&\bar{\E}_{j\ell}&=1,\quad&\bar{\E}_{jm}&=1,\quad&\bar{\E}_{k\ell}&=1,\quad&\bar{\E}_{km}&=1,\quad&\bar{\E}_{\ell m}&=1,\\
\E_{jk\ell}&=1,\quad&\E_{jkm}&=1,\quad&\E_{j\ell m}&=1,\quad&\E_{k\ell m}&=1
\end{alignedat}
\intertext{and that the latter system is equivalent to}
&x_{\ell m}x_{jkm}-x_{km}x_{j\ell m}+x_{jm}x_{k\ell m}=0,\\
&x_{km}x_{jk\ell}-x_{k\ell}x_{jkm}+x_{jk}x_{k\ell m}=0,\\
&x_{\ell m}x_{jk\ell}-x_{k\ell}x_{j\ell m}+x_{j\ell}x_{k\ell m}=0,\\
&x_{jm}x_{jk\ell}-x_{j\ell}x_{jkm}+x_{jk}x_{j\ell m}=0,\\
&x_{jk}x_{\ell m}-x_{j\ell}x_{km}+x_{jm}x_{k\ell}=0.
\end{align*}
On the other hand, if we consider a solution $x$ of~\eqref{eq:EL cube} such that $\E_{j}=1$ and $\E_{k}=1$, we know from the proof of Theorem~\ref{th:corner white} that
\begin{align*}
&\begin{alignedat}{6}
\bar{\E}_{jk}&=1,\quad&\bar{\E}_{j\ell}&=1,\quad&\bar{\E}_{jm}&=1,\quad&\bar{\E}_{k\ell}&=1,\quad&\bar{\E}_{km}&=1,\quad&\bar{\E}_{\ell m}&=1,\\
\E_{jk\ell}&=1,\quad&\E_{jkm}&=1,\quad&\E_{j\ell m}&=1,\quad&\E_{k\ell m}&=1
\end{alignedat}
\intertext{and that the latter system is equivalent to}
&x_{km}x_{\ell m}x_{jkm}x_{j\ell m}-x_{jm}x_{\ell m}x_{jkm}x_{k\ell m}+x_{jm}x_{km}x_{j\ell m}x_{k\ell m}=0,\\
&x_{k\ell}x_{\ell m}x_{jk\ell}x_{j\ell m}-x_{j\ell}x_{\ell m}x_{jk\ell}x_{k\ell m}+x_{j\ell}x_{k\ell}x_{j\ell m}x_{k\ell m}=0,\\
&x_{k\ell}x_{km}x_{jk\ell}x_{jkm}-x_{jk}x_{km}x_{jk\ell}x_{k\ell m}+x_{jk}x_{k\ell}x_{jkm}x_{k\ell m}=0,\\
&x_{j\ell}x_{jm}x_{jk\ell}x_{jkm}-x_{jk}x_{jm}x_{jk\ell}x_{j\ell m}+x_{jk}x_{j\ell}x_{jkm}x_{j\ell m}=0,\\
&x_{j\ell}x_{jm}x_{k\ell}x_{km}-x_{jk}x_{jm}x_{k\ell}x_{\ell m}+x_{jk}x_{j\ell}x_{km}x_{\ell m}=0.
\end{align*}
Since a solution $x$ of~\eqref{eq:EL cube} cannot solve
\begin{align*}
&x_{jk}x_{\ell m}-x_{j\ell}x_{km}+x_{jm}x_{k\ell}=0
\intertext{and}
&x_{j\ell}x_{jm}x_{k\ell}x_{km}-x_{jk}x_{jm}x_{k\ell}x_{\ell m}+x_{jk}x_{j\ell}x_{km}x_{\ell m}=0
\end{align*}
at the same time, this proves the theorem.
\end{proof}
\end{theo}
\begin{theo}[Closure relation]
There holds $S^{jk\ell m}=0$ on all solutions of~\eqref{eq:EL cube}.
\begin{proof}
Let $x$ be a solution of~\eqref{eq:DKP case cube} or~\eqref{eq:inverse DKP case cube}. Then
\[
S^{jk\ell m}=d\ELL(\{jk\ell\})=d\Ell(P_{i}\lfloor ijk\ell m\rfloor)+d\Ell(-P_{i}T_{\bar{\imath}}\lceil ijk\ell m\rceil)=\unbar{S}^{ijk\ell m}-\bar{S}^{ijk\ell m}=\pm\frac{\pi^{2}}{4}\mp\frac{\pi^{2}}{4}=0
\]
due to Theorems~\ref{th:closure black} and \ref{th:closure white} since every solution of~\eqref{eq:DKP case cube} solves~\eqref{eq:DKP case} and~\eqref{eq:DKP case white} after the transformation $P_{i}$ of variables and every solution of~\eqref{eq:inverse DKP case cube} solves~\eqref{eq:inverse DKP case} and~\eqref{eq:inverse DKP case white} after the transformation $P_{i}$ of variables.
\end{proof}
\end{theo}

\section{Conclusion}
The fact that the three-dimensional (hyperbolic) dKP equation is, in a sense, equivalent to the Euler-Lagrange equations of the corresponding action is rather surprising since for the two-dimensional (hyperbolic) quad-equations an analogous statement is not true (see~\cite{variational,octahedron} for more details). On the other hand, in the continuous situation there is an example of a 2-form whose Euler-Lagrange equations are equivalent to the set of equations consisting of the (hyperbolic) sine-Gordon equation and the (evolutionary) modified Korteweg-de~Vries equation (see~\cite{S13} for more details). So, the general picture remains unclear.\par
In particular, the variational formulation for the other equations of octahedron type in the classification of \cite{ABS3} is still an open problem.

\section*{Acknowledgments}
This research was supported by the DFG Collaborative Research Center TRR 109 ``Discretization in Geometry and Dynamics''.

\appendix
\section{Facets of \texorpdfstring{$N$}{N}-cells of the root lattice \texorpdfstring{$Q(A_{N})$}{Q(A\_N)}}\label{sec:facets}
\paragraph{Facets of 3-cells:}\ \\
\begin{tabular}{ll}\addlinespace
Black tetrahedra $\lfloor ijk\ell\rfloor$:& four black triangles $\lfloor ijk\rfloor$, $-\lfloor ij\ell\rfloor$, $\lfloor ik\ell\rfloor$, and $-\lfloor jk\ell\rfloor$;\\\addlinespace
Octahedra $[ijk\ell]$:& four black triangles $T_{\ell}\lfloor ijk\rfloor$, $-T_{k}\lfloor ij\ell\rfloor$, $T_{j}\lfloor ik\ell\rfloor$, and $-T_{i}\lfloor jk\ell\rfloor$,\\
& four white triangles $\lceil ijk\rceil$, $-\lceil ij\ell\rceil$, $\lceil ik\ell\rceil$, and $-\lceil jk\ell\rceil$;\\\addlinespace
White tetrahedra $\lceil ijk\ell\rceil$:& four white triangles $T_{\ell}\lceil ijk\rceil$, $-T_{k}\lceil ij\ell\rceil$, $T_{j}\lceil ik\ell\rceil$, and $-T_{i}\lceil jk\ell\rceil$;
\end{tabular}
\paragraph{Facets of 4-cells:}\ \\
\begin{tabular}{ll}\addlinespace
Black 4-simplices $\llfloor ijk\ell m\rrfloor$:& five black tetrahedra $\lfloor ijk\ell\rfloor$, $-\lfloor ijkm\rfloor$, $\lfloor ij\ell m\rfloor$,\\
& $-\lfloor ik\ell m\rfloor$, and $\lfloor jk\ell m\rfloor$;\\\addlinespace
Black 4-ambo-simplices $\lfloor ijk\ell m\rfloor$:& five black tetrahedra $T_{m}\lfloor ijk\ell\rfloor$, $-T_{\ell}\lfloor ijkm\rfloor$, $T_{k}\lfloor ij\ell m\rfloor$,\\
& $-T_{j}\lfloor ik\ell m\rfloor$, and $T_{i}\lfloor jk\ell m\rfloor$,\\
& and five octahedra $[ijk\ell]$, $-[ijkm]$, $[ij\ell m]$, $-[ik\ell m]$,\\
& and $[jk\ell m]$;
\\\addlinespace
White 4-ambo-simplices $\lceil ijk\ell m\rceil$:& five octahedra $T_{m}[ijk\ell]$, $-T_{\ell}[ijkm]$, $T_{k}[ij\ell m]$,\\
& $-T_{j}[ik\ell m]$, and $T_{i}[jk\ell m]$,\\
& and five white tetrahedra $\lceil ijk\ell\rceil$, $-\lceil ijkm\rceil$, $\lceil ij\ell m\rceil$,\\
& $-\lceil ik\ell m\rceil$, and $\lceil jk\ell m\rceil$;
\end{tabular}\par
\begin{tabular}{ll}
White 4-simplices $\llceil ijk\ell m\rrceil$:& five white tetrahedra $T_{m}\lceil ijk\ell\rceil$, $-T_{\ell}\lceil ijkm\rceil$,\\
& $T_{k}\lceil ij\ell m\rceil$, $-T_{j}\lceil ik\ell m\rceil$, and $T_{i}\lceil jk\ell m\rceil$.
\end{tabular}
\section{4D corners on 4-cells of the root lattice \texorpdfstring{$Q(A_{N})$}{Q(A\_N)}}\label{sec:corners}
\paragraph{Black 4-simplex $\llfloor ijk\ell m\rrfloor$:} The 4D corner with center vertex $x_{i}$ contains
\begin{itemize}
\item the four black tetrahedra $\lfloor ijk\ell\rfloor$, $-\lfloor ijkm\rfloor$, $\lfloor ij\ell m\rfloor$, and $-\lfloor ik\ell m\rfloor$;
\end{itemize}
\paragraph{Black 4-ambo-simplex $\lfloor ijk\ell m\rfloor$:} The 4D~corner with center vertex $x_{ij}$ contains
\begin{itemize}
\item the two black tetrahedra $-T_{j}\lfloor ik\ell m\rfloor$, and $T_{i}\lfloor jk\ell m\rfloor$,
\item and the three octahedra $[ijk\ell]$, $-[ijkm]$, and $[ij\ell m]$;
\end{itemize}
\paragraph{White 4-ambo-simplex $\lceil ijk\ell m\rceil$:} The 4D~corner with center vertex $x_{ijk}$ contains
\begin{itemize}
\item the three octahedra $T_{k}[ij\ell m]$, $-T_{j}[ik\ell m]$, and $T_{i}[jk\ell m]$,
\item and the two white tetrahedra $\lceil ijk\ell\rceil$, and $-\lceil ijkm\rceil$;
\end{itemize}
\paragraph{White 4-simplex $\llceil ijk\ell m\rrceil$:} The 4D~corner with center vertex $x_{ijk\ell}$ contains
\begin{itemize}
\item the four white tetrahedra $-T_{\ell}\lceil ijkm\rceil$, $T_{k}\lceil ij\ell m\rceil$, $-T_{j}\lceil ik\ell m\rceil$, and $T_{i}\lceil jk\ell m\rceil$.
\end{itemize}

\section{Proof of Theorem~\ref{th:flower}}\label{sec:proof}
Set $M:=N+1$ and $L:=N+2$. Then, for the construction of the sum $\Sigma$ of 4D~corners representing the flower $\sigma$ centered in $X$, we use the following algorithm:
\begin{enumerate}
\item For every black tetrahedron $\pm\lfloor ijk\ell\rfloor\in\sigma$ at the interior vertex $X$ we add the 4D~corner with center vertex $X$ on the black 4-simplex $\pm\llfloor ijk\ell M\rrfloor$ to $\Sigma$.
\item For every octahedron $\pm[ijk\ell]\in\sigma$ we add the 4D~corner with center vertex $X$ on the black 4-ambo-simplex $\pm\lfloor ijk\ell M\rfloor$ to $\Sigma$.
\item For every white tetrahedron $\pm\lceil ijk\ell\rceil\in\sigma$ we add the 4D~corner with center vertex $X$ on the white 4-ambo-simplex $\pm\lceil ijk\ell M\rceil$ to $\Sigma$.
\item For every white tetrahedron $\pm\lceil ijkM\rceil\in\sigma$ which appeared in $\Sigma$ during the previous step we add the 4D~corner with center vertex $X$ on the white 4-simplex $\mp T_{\bar{L}}\llceil ijkML\rrceil$ to $\Sigma$.
\end{enumerate}
Therefore, we have to prove that $\Sigma=\sigma$.\par
Assume that $X=x_{i}$. Then for each black tetrahedron $\pm\lfloor ijk\ell\rfloor\in\sigma$ we added the three black tetrahedra $\mp\lfloor ijkM\rfloor$, $\pm\lfloor ij\ell M\rfloor$, and $\mp\lfloor ik\ell M\rfloor$ to $\Sigma$ which do not belong to $\sigma$. Moreover, $\pm\lfloor ijk\ell\rfloor$ has three black triangular facets adjacent to $x_{i}$, namely $\pm\lfloor ijk\rfloor$, which is the common triangle with $\mp\lfloor ijkM\rfloor$, $\mp\lfloor ij\ell\rfloor$ (up to orientation), which is the common triangle with $\pm\lfloor ij\ell M\rfloor$, and $\pm\lfloor ik\ell\rfloor$, which is the common triangle with $\mp\lfloor ik\ell M\rfloor$. Therefore, each of these black tetrahedra has to cancel away with the corresponding black tetrahedra from the 4D~corner which is coming from the 3-cell adjacent to $\pm\lfloor ijk\ell\rfloor$ via the corresponding black triangle.\par
Assume that $X=x_{ij}$. Then for each octahedron $\pm[ijk\ell]\in\sigma$ we added the two black tetrahedra $\mp T_{j}\lfloor ik\ell M\rfloor$ and $\pm T_{i}\lfloor jk\ell M\rfloor$ as well as the two octahedra $\mp[ijkM]$ and $\pm[ij\ell M]$ to $\Sigma$ which do not belong to $\sigma$. Moreover, $\pm[ijk\ell]$ has two black tetrahedral facets adjacent to $x_{ij}$, namely $\pm\lfloor ijk\rfloor$, which is the common triangle with $\mp T_{j}\lfloor ik\ell M\rfloor$, and $\mp\lfloor ij\ell\rfloor$, which is the common triangle with $\pm T_{i}\lfloor jk\ell M$, as well as two white tetrahedral facets adjacent to $x_{ij}$, namely $\pm T_{j}\lceil ik\ell\rceil$, which is the common triangle with $\mp[ijkM]$ and $\pm[ij\ell M]$, and $\mp\lceil jk\ell\rceil$, which is the common triangle with $\pm[ij\ell M]$. Therefore, each of the black tetrahedra $\mp T_{j}\lfloor ik\ell M\rfloor$ and $\pm T_{i}\lfloor jk\ell M$ has to cancel away with the corresponding black tetrahedron from the 4D~corner which is coming from the 3-cell adjacent to$\pm[ijk\ell]$ via the corresponding black triangle, and each of the octahedra $\mp[ijkM]$ and $\pm[ij\ell M]$ has to cancel away with the corresponding octahedron coming the 4D~corner which is coming from the 3-cell adjacent to $\pm[ijk\ell]$ via the corresponding white triangle.\par
Assume that $X=x_{ijk}$. Then for each white tetrahedron $\pm\lceil ijk\ell\rceil\in\sigma$ we added the three octahedra $\pm T_{k}[ij\ell M]$, $\mp T_{j}[ik\ell M]$, and $\pm T_{i}[jk\ell M]$ as well as the white tetrahedron $\mp\lceil ijkM\rceil$ to $\Sigma$ which do not belong to $\sigma$. Moreover, $\pm\lceil ijk\ell\rceil$ has three white triangular facets adjacent to $x_{ijk}$, namely $\mp T_{k}\lceil ij\ell\rceil$, which is the common triangle with $\pm T_{k}[ij\ell M]$, $\pm T_{j}\lceil ik\ell\rceil$, which is the common triangle with $\mp T_{j}[ik\ell M]$, and $\mp T_{i}\lceil jk\ell\rceil$, which is the common triangle with $\pm T_{i}[jk\ell M]$. Therefore, each of these octahedra has to cancel away with the corresponding octahedron from the 4D~corner which is coming from the 3-cell adjacent to $\pm\lceil ijk\ell\rceil$ via the corresponding white triangle.\par
Consider two 3-cells $\Omega,\bar{\Omega}\in\sigma$ adjacent via the black triangle $\lfloor ijk\rfloor$, say $\lfloor ijk\rfloor$ belongs to $\Omega$ and $-\lfloor ijk\rfloor$ belongs to $\bar{\Omega}$. Then the 4D~corner corresponding to $\Omega$ contributes the black tetrahedron $-\lfloor ijkM\rfloor$ to $\Sigma$, whereas the 4D~corner corresponding to $\bar{\Omega}$ contributes the black tetrahedron $\lfloor ijkM\rfloor$ to $\Sigma$. Therefore, the latter two black tetrahedra cancel out.\par
Consider two 3-cells $\Omega,\bar{\Omega}\in\sigma$ adjacent via the white triangle $\lceil ijk\rceil$, say $\lceil ijk\rceil$ belongs to $\Omega$ and $-\lceil ijk\rceil$ belongs to $\bar{\Omega}$. Then the 4D~corner corresponding to $\Omega$ contributes the octahedron $-[ijkM]$ to $\Sigma$, whereas the 4D~corner corresponding to $\bar{\Omega}$ contributes the octahedron $[ijkM]$ to $\Sigma$. Therefore, the latter two octahedra cancel out.\par
Up to know we proved that all black tetrahedra and all octahedra in $\Sigma\setminus\sigma$ cancel out. We will now consider with the white tetrahedra in $\Sigma\setminus\sigma$.\par
\begin{lemma}
The white tetrahedra $\lfloor ijk M\rfloor$ arising in the third step of the algorithm build flowers which only contain white tetrahedra.
\begin{proof}
We have two prove that each of these white tetrahedra has exactly three adjacent white tetrahedra in the flowers.\par
Assume that $X=x_{ij}$ and consider the two adjacent white tetrahedra $\pm T_{\bar{k}}\lceil ijk\ell\rceil$ and $\pm T_{\bar{m}}\lceil ij\ell m\rceil$ in $\sigma$. For these white tetrahedra we added -- during the third step of the algorithm -- the two 4D~corners with center vertex $x_{ij}$ on the white 4-ambo-simplices $\pm T_{\bar{k}}\lceil ijk\ell M\rceil$ and $\pm T_{\bar{m}}\lceil ij\ell mM\rceil$ to $\Sigma$. These two flowers contain exactly two tetrahedra which are not in $\sigma$, namely $\mp T_{\bar{k}}\lceil ijkM\rceil$ and $\pm T_{\bar{m}}\lceil ijmM\rceil$, which are adjacent to each other via the white triangle $\lceil ijM\rceil$.\par
Consider an octahedron $\pm[ijk\ell]\in\sigma$ and assume that $X=x_{ij}$. It has exactly two adjacent 3-cells via white triangles. Therefore, one can say that octahedra appear only in chains, either in closed chains $\pm[ijk_{1}k_{2}],\pm[ijk_{2}k_{3}],\ldots,\pm[ijk_{\alpha}k_{1}]$ with $\alpha\in\N\setminus\{0,1\}$ or in open chains
$\pm T_{\bar{k}}\lceil ijk\ell_{1}\rceil, \pm[ij\ell_{1}\ell_{2}],\pm[ij\ell_{2}\ell_{3}],\ldots,\pm[ijk_{\alpha-1}k_{\alpha}],\pm T_{\bar{m}}\lceil ij\ell_{\alpha}m\rceil$ with $\alpha\in\N\setminus\{0,1\}$, where the first and the last octahedron are adjacent two white tetrahedra. Here, it may happen that the letters in the brackets are not increasingly ordered, but this does not affect the result. Since octahedra in $\sigma$ do not lead to white tetrahedra in $\Sigma$, we are only interested in open chains. Moreover, we only consider the ``$+$''-case. The ``$-$''-case is analogous.\par
For the white tetrahedra $T_{\bar{k}}\lceil ijk\ell_{1}\rceil$ and $T_{\bar{m}}\lceil ij\ell_{\alpha}m\rceil$ we added -- in the third step of the algorithm -- the 4D~corners  with center vertex $x_{ij}$ on the white 4-ambo-simplices $T_{\bar{k}}\lceil ijk\ell_{1}M\rceil$ and $T_{\bar{m}}\lceil ij\ell_{\alpha}mM\rceil$ to $\Sigma$. These two 4D~corners contain exactly two white tetrahedra which do not belong to $\sigma$, namely $-T_{\bar{k}}\lceil ijkM\rceil$ and $T_{\bar{m}}\lceil ijmM\rceil$. Now, we have to consider two cases:
\begin{itemize}
\item $k\neq m$, i.e., $T_{\bar{k}}\lceil ijk\ell_{1}\rceil$ and $T_{\bar{m}}\lceil ij\ell_{\alpha}m\rceil$ do not belong to a common 4-ambo-simplex: here, $-T_{\bar{k}}\lceil ijkM\rceil$ and $T_{\bar{m}}\lceil ijmM\rceil$ are adjacent to each other via the white triangle $\lceil ijM\rceil$. Comparing this result with the previous one about two adjacent white tetrahedra in $\sigma$, we realize that it makes no difference for the resulting tetrahedra whether the original tetrahedra are adjacent or connected by a chain of octahedra as long as the do not belong to a common 4-ambo-simplex.
\item $k=m$, i.e., $T_{\bar{k}}\lceil ijk\ell_{1}\rceil$ and $T_{\bar{m}}\lceil ij\ell_{\alpha}m\rceil$ both belong to the 4-ambo-simplex $T_{\bar{k}}\lceil ijk\ell_{1}\ell_{\alpha}\rceil$: here, $-T_{\bar{k}}\lceil ijkM\rceil$ and $T_{\bar{m}}\lceil ijmM\rceil$ cancel out. Therefore, we have to prove that other white tetrahedra which are adjacent to one of these two white tetrahedra have exactly three adjacent white tetrahedra in the flowers. Due to the remark in the previous case we can -- without loss of generality -- assume that $\sigma$ contains the white tetrahedron $T_{i}T_{\bar{k}}T_{\bar{n}}\lceil jk\ell_{1} n\rceil$ which is adjacent to $T_{\bar{k}}\lceil ijk\ell_{1}\rceil$ via the white triangle $T_{i}T_{\bar{k}}\lceil jk\ell_{1}\rceil$. Therefore, it turns out that $T_{\bar{k}}\lceil ijk\ell_{1}\rceil$ and $-T_{\bar{k}}\lceil ijk\ell_{\alpha}\rceil$ cannot be connected by the chain $T_{i}T_{\bar{k}}[jk\ell_{1}\ell_{\alpha+1}],T_{i}T_{\bar{k}}[jk\ell_{\alpha+1}\ell_{\alpha+2}],\ldots,T_{i}T_{\bar{k}}[jk\ell_{\beta}\ell_{\alpha}]$ with $\beta\in\N$, $\beta>\alpha$, and we can assume that $\sigma$ contains the white tetrahedron $-T_{i}T_{\bar{k}}T_{\bar{p}}\lceil jk\ell_{\alpha}p\rceil$ which is adjacent to $-T_{\bar{k}}\lceil ijk\ell_{\alpha}\rceil$ via the white triangle $T_{i}T_{\bar{k}}\lceil jk\ell_{\alpha}\rceil$. For the white tetrahedra $T_{i}T_{\bar{k}}T_{\bar{n}}\lceil jk\ell_{1} n\rceil$ and $-T_{i}T_{\bar{k}}T_{\bar{p}}\lceil jk\ell_{\alpha} p\rceil$ we added -- in the third step of the algorithm -- the 4D~corners with center vertex $x_{ij}$ on the white 4-ambo-simplices $T_{i}T_{\bar{k}}T_{\bar{n}}\lceil jk\ell_{1} nM\rceil$ and $-T_{i}T_{\bar{k}}T_{\bar{p}}\lceil jk\ell_{\alpha}pM\rceil$ to $\Sigma$. These two 4D~corners contain exactly two white tetrahedra which do not belong to $\sigma$, namely $T_{i}T_{\bar{k}}T_{\bar{n}}\lceil jknM\rceil$ and $-T_{i}T_{\bar{k}}T_{\bar{p}}\lceil jkpM\rceil$ which are adjacent via the white triangle $T_{i}T_{\bar{k}}\lceil jkM\rceil$.\qedhere
\end{itemize}
\end{proof}
\end{lemma}
Now we continue with the proof of Theorem~\ref{th:flower}. We assume that $X=x_{ij}$ and consider the white tetrahedron $T_{\bar{k}}\lceil ijkM\rceil\in\Sigma$ in the flowers which appeared in the third step of the algorithm. For this white tetrahedron we added -- in the fourth step of the algorithm -- the 4D~corner with center vertex $x_{ij}$ on the white 4-simplex $-T_{\bar{k}}T_{\bar{L}}\llceil ijkML\rrceil$ to $\Sigma$. This 4D~corner contains the four white tetrahedra
$-T_{\bar{k}}\lceil ijkM\rceil$, $-T_{\bar{L}}\lceil ijML\rceil$, $T_{j}T_{\bar{k}}T_{\bar{L}}\lceil ikML\rceil$, and $-T_{i}T_{\bar{k}}T_{\bar{L}}\lceil jkML\rceil$. Therefore, the white tetrahedra $T_{\bar{k}}\lceil ijkM\rceil$ and $-T_{\bar{k}}\lceil ijkM\rceil$ cancel out in $\Sigma$. Furthermore, we consider the white tetrahedron $-T_{\bar{m}}\lceil ijmM\rceil\in\Sigma$ which also appeared in the third step of the algorithm and is adjacent to the white tetrahedron $T_{\bar{k}}\lceil ijkM\rceil$ via the white triangle $\lceil ijM\rceil$. For this white tetrahedron we added -- in the fourth step of the algorithm -- the 4D~corner with center vertex $x_{ij}$ on the white 4-simplex $T_{\bar{m}}T_{\bar{L}}\llceil ijmML\rrceil$ to $\Sigma$. This 4D~corner contains the for white tetrahedra
$T_{\bar{m}}\lceil ijmM\rceil$, $T_{\bar{L}}\lceil ijML\rceil$, $-T_{j}T_{\bar{k}}T_{\bar{L}}\lceil imML\rceil$, and $T_{i}T_{\bar{k}}T_{\bar{L}}\lceil jmML\rceil$. Therefore, the white tetrahedra $-T_{\bar{m}}\lceil ijmM\rceil$ and $T_{\bar{m}}\lceil ijmM\rceil$ as well as the white tetrahedra $-T_{\bar{L}}\lceil ijML\rceil$ and $T_{\bar{L}}\lceil ijML\rceil$ cancel out in $\Sigma$.\qed

{\small
\bibliographystyle{amsalpha}
\providecommand{\bysame}{\leavevmode\hbox to3em{\hrulefill}\thinspace}
\providecommand{\MR}{\relax\ifhmode\unskip\space\fi MR }
\providecommand{\MRhref}[2]{%
  \href{http://www.ams.org/mathscinet-getitem?mr=#1}{#2}
}
\providecommand{\href}[2]{#2}

}

\end{document}